\documentclass{aa}

\usepackage{graphicx} 
\usepackage{bm}       
\usepackage{txfonts}
\usepackage{natbib}

\def\cgfit{\textsc{gFIT}}
\def\gfit{\textsc{gFIT}}
\def\galsim{\textsc{Galsim}}
\def\pse{\textsc{SExtractor}}
\def\ksb{\textsc{KSB}}
\def\metacal{\textsc{MetaCalibration}}
\def\shapelens{\textsc{shapelens}}

\newcommand{\mat}[1]{\bm{\mathrm{#1}}}

\newcommand{\Fref}[1]{Fig.~\ref{#1}}

\begin{document}

\title{
Shear measurement bias II: a fast machine-learning calibration method}
\titlerunning{Calibration methods}

\authorrunning
{Arnau Pujol\and
Jerome Bobin\and
Florent Sureau\and
et al.
}
\author
{Arnau Pujol \inst{1,2,3,4} \and
Jerome Bobin \inst{1,2,5} \and
Florent Sureau \inst{1,2} \and
Axel Guinot \inst{1,2} \and
Martin Kilbinger \inst{1,2, 6} \\
}
\institute{
DEDIP/DAP, IRFU, CEA, Universit\'e Paris-Saclay, F-91191 Gif-sur-Yvette, France
\and
AIM, CEA, CNRS, Université Paris-Saclay, Université Paris Diderot, Sorbonne Paris Cité, F-91191 Gif-sur-Yvette, France
\and
Institut d’ Estudis Espacials de Catalunya (IEEC), E-08034 Barcelona, Spain
\and
Institute of Space Sciences (ICE, CSIC), E-08193 Barcelona, Spain
\and
Institute of Particle and Cosmos Physics (IPARCOS), Universidad Complutense de Madrid, E-28040 Madrid, Spain
\and
Institut d'Astrophysique de Paris, UMR7095 CNRS, Universit\'e Pierre \& Marie Curie, 98 bis boulevard Arago, F-75014 Paris, France\\
}

\date{Received date / Accepted date}

\abstract{
We present a new shear calibration method based on machine learning. The method estimates the individual shear responses of the objects from the combination of several measured properties on the images using supervised learning. The supervised learning uses the true individual shear responses obtained from copies of the image simulations with different shear values. On simulated GREAT3 data, we obtain a residual bias after the calibration compatible with $0$ and beyond Euclid requirements for a signal-to-noise ratio $>20$ within $\sim 15$ CPU hours of training using only $\sim 10^5$ objects. This efficient machine-learning approach can use a smaller data set because the method avoids the contribution from shape noise. The low dimensionality of the input data also leads to simple neural network architectures.  We compare it to the recently described method Metacalibration, which shows similar performances. The different methods and systematics suggest that the two methods are very good complementary methods. Our method can therefore be applied without much effort to any survey such as Euclid or the Vera C. Rubin Observatory, with fewer than a million images to simulate to learn the calibration function.}


\maketitle

\section{Introduction}

Weak gravitational lensing by the large-scale structure has become an
important tool for cosmology in recent years. Light deflection by tidal fields
of the inhomogeneous matter on very large scales causes small deformations of
images of high-redshift galaxies. This cosmic shear contains valuable
information about the growth of structures in the Universe and can help to
shed light on the nature of dark matter and dark energy. The amount of shear
that is induced by weak lensing is very small, at the percent level, and should be estimated based on high-accuracy galaxy images for a reliable cosmological inference: measurement biases need to be reduced to the sub-percent
level to pass the requirement of upcoming large cosmic-shear experiments, such
as the ESA space mission Euclid \citep{Laureijs2011}, the NASA space satellite
Roman Space Telescope \citep{2019arXiv190205569A}, or the ground-based Vera C. Rubin Observatory, previously referred to as the Large Synoptic Survey Telescope \citep{LSST2009}.

Shear is estimated by measuring galaxy shapes and averaging out their
intrinsic ellipticity. This estimate is in general biased by noise,
inappropriate assumptions about the galaxy light distribution, uncorrected point spread function (PSF)
residuals, and detector effects such as the brighter-fatter effect or the charge
transfer inefficiency
\citep{Bridle2009,Bridle2010,Kitching2010,Kitching2012,Kitching2013,Refregier2012,Kacprzak2012,Melchior2012,Taylor2016,Massey2007b,Massey2013,Voigt2010,Bernstein2010,Zhang2011,Kacprzak2012,Kacprzak2014,Mandelbaum2015,Clampitt2017}.
The resulting shear biases are complex functions of many
parameters that describe galaxy and instrument properties. These include the galaxy
size, flux, morphology, signal-to-noise ratio (S/N), intrinsic ellipticity, PSF size,
anisotropy and its alignment with respect to the galaxy orientation, and many
more \citep{Zuntz2013,Conti2016,Hoekstra2015,Hoekstra2017,Pujol2017}.


To achieve the sub-percent shear bias that is expected in future cosmic-shear surveys, the shear estimates typically need to be calibrated using a very high number of simulated images, for instance to overcome the statistical variability induced by galaxy intrinsic shapes \citep{Massey2013}. Furthermore, these simulations need to adequately span the high-dimensional space of parameters that determines the shear bias. Otherwise, regions of parameter space that are underrepresented in the simulations compared to the observations can lead to incorrect bias correction. The selection of objects needs to closely match the real selection function to avoid selection biases.

Existing calibration methods requiring extensive image simulations select a few parameters of interest a priori,
often galaxy size and S/N, for which
the shear bias variation is estimated \citep{Zuntz2017}. The shear bias is computed using various methods such as
fitting to the parameters \citep{Jarvis2015,2018PASJ...70S..25M} or
$k$-nearest neighbours \citep{Hildebrandt2016}.

Machine-learning techniques have also been employed for shear
estimation and calibration.
\cite{2010ApJ...720..639G} trained an artificial neural network (NN) to minimise the shear bias
from parameters measured in the moment-based method KSB \citep{Kaiser1995}.
More recently, \cite{Tewes2018} presented an artificial neural
network for supervised learning to obtain shear estimates from a few fitted parameters from images using adaptive weighted moments via regression and using image simulations with varying galaxy features as a training set.

An alternative shear calibration method that does not require image simulations
and is based on the data themselves is the so-called meta-calibration
\citep{Huff2017}. This approach computes the shear response matrix by adding
low shear values to deconvolved observed galaxy images. A hybrid method is
a self-calibration \citep{Conti2016}, for which noise-free simulated images are
created and re-measured, according to the best-fit parameters measured
on the data, to reduce noise bias.




This paper extends previous work of machine learning for shear calibration. In the companion paper,
\cite{Pujol2017}, hereafter Paper~I, we have explored the dependence of shear bias
 on various combinations of input and measured parameters. We demonstrated the complexity
of this shear bias function and showed that it is important to account for
correlations between parameters. A multi-dimensional parameter space of galaxy and PSF properties is therefore set up to learn the shear bias function using a deep-learning architecture to regress the shear bias from these parameters.


%
This paper is organized as follows.
Sect.~\ref{sec:shear_bias_measurement}
presents the definition of shear bias and a review of our method for measuring shear bias that was
introduced in \cite{Pujol2018} (hereafter PKSB19).
In Sect.~\ref{sec:calibration} we introduce our new shear calibration method.
Sect.~\ref{sec:data} presents the simulated images and the input data used for
the training, testing, and validation of our method to produce the results of this paper,
 which are discussed and compared to an existing method in Sect.~\ref{sec:results}. After a
 discussion of several points regarding the new method in Sect.~\ref{sec:discussion},
we conclude with a summary of the study in Sect.~\ref{sec:conclusions}.

\section{Shear bias}\label{sec:shear_bias_measurement}

\subsection{Shear bias definition}

In the weak-lensing regime, the observed ellipticity of a galaxy $e_i^\textrm{obs}$ is an estimator of the reduced shear $g_i$ for component $i=1, 2$. In general, however, this estimator is biased by pixel noise, PSF residuals, inaccurate galaxy models, and other effects (see \cite{Mandelbaum2018} for a recent review). The bias of the estimated shear, $g^\textrm{obs}$, is usually expressed by the following equation:
\begin{equation}
\langle e_i^{\rm obs} \rangle =  g_i^{\rm obs} = c_i + (1 + m_i) g_i ,
\label{eq:g_relation}
\end{equation}
where $c_{i}\text{ and }m_{i}$ are the additive and multiplicative shear biases, respectively.
For a constant shear, if the mean intrinsic ellipticity of a galaxy sample is zero, we can measure the shear from the average observed ellipticities using the above relation.

We can also define the response of the ellipticity measurement of an image to linear changes in the shear \citep{Huff2017},
\begin{equation}
R_{ij}  =  \frac{\partial e_i^{\rm obs}}{\partial g_j}.
\label{eq:shear_response}
\end{equation}

The multiplicative bias of a population can be obtained from the average shear responses, as described in \cite{Pujol2018},
\begin{equation}
1 + m_i = \langle R_{ii} \rangle.
\label{eq:mean_response}
\end{equation}
In a similar way, the additive population shear bias can be obtained from the average of the individual additive biases,
\begin{equation}
c_i = \langle e_i^{\rm obs} - e_i^{\rm I} - R_{ii}g_i \rangle = \langle e_i^{\rm obs} \rangle,
\label{eq:ind_add_bias}
\end{equation}
where $e_i^{\rm I}$ is the intrinsic ellipticity, and the second equality holds if $\langle e_i^{\rm I} \rangle = 0$ and $\langle R_{ii}g_i \rangle = 0$ (which is true if $R_{ii}$ is not correlated with $g_i$ and $\langle g_i \rangle = 0$).

\subsection{Shear bias measurements}\label{sec:shear_measurements}

We used two methods to estimate shear bias, both of which we briefly describe in this section.
For more details on the different methods for estimating shear bias in simulations we refer to PKSB19.

First, to test the residual shear bias after calibration, we used the common approach to measure shear bias from a linear fit to eq.~(\ref{eq:g_relation}).
We simulated each galaxy with its orthogonal pair to guarantee that the average intrinsic ellipticity is zero. This improves the precision of the shear bias estimation by a factor of $\sim 3$, as shown in PKSB19.

Second, to study the shear bias dependences, we used the method introduced in PKSB19: We measured the individual shear responses and additive biases from each galaxy image by using different sheared versions of the same image. The multiplicative and additive bias of a population was then obtained from the average of these quantities. This method has been proven to be more precise by a factor of $\sim 12$ than the linear fit with orthogonal-pair noise cancellation (PKSB19).

These individual shear responses and additive biases are also used as
quantities for the supervised machine-learning algorithm of our calibration
method (Sect.~\ref{sec:learning_description}). We denote the biases
measured from simulations as described here `true' biases $m_i^{\textrm{t}},
c_i^{\textrm{t}}$, indicated with the superscript `t`. Our goal in this paper
is to regress these biases in a high-dimensional parameter space where this function is living.

\subsection{Dependences}\label{sec:dependencies}

\begin{figure}
\centering
\includegraphics[width=.95\linewidth]{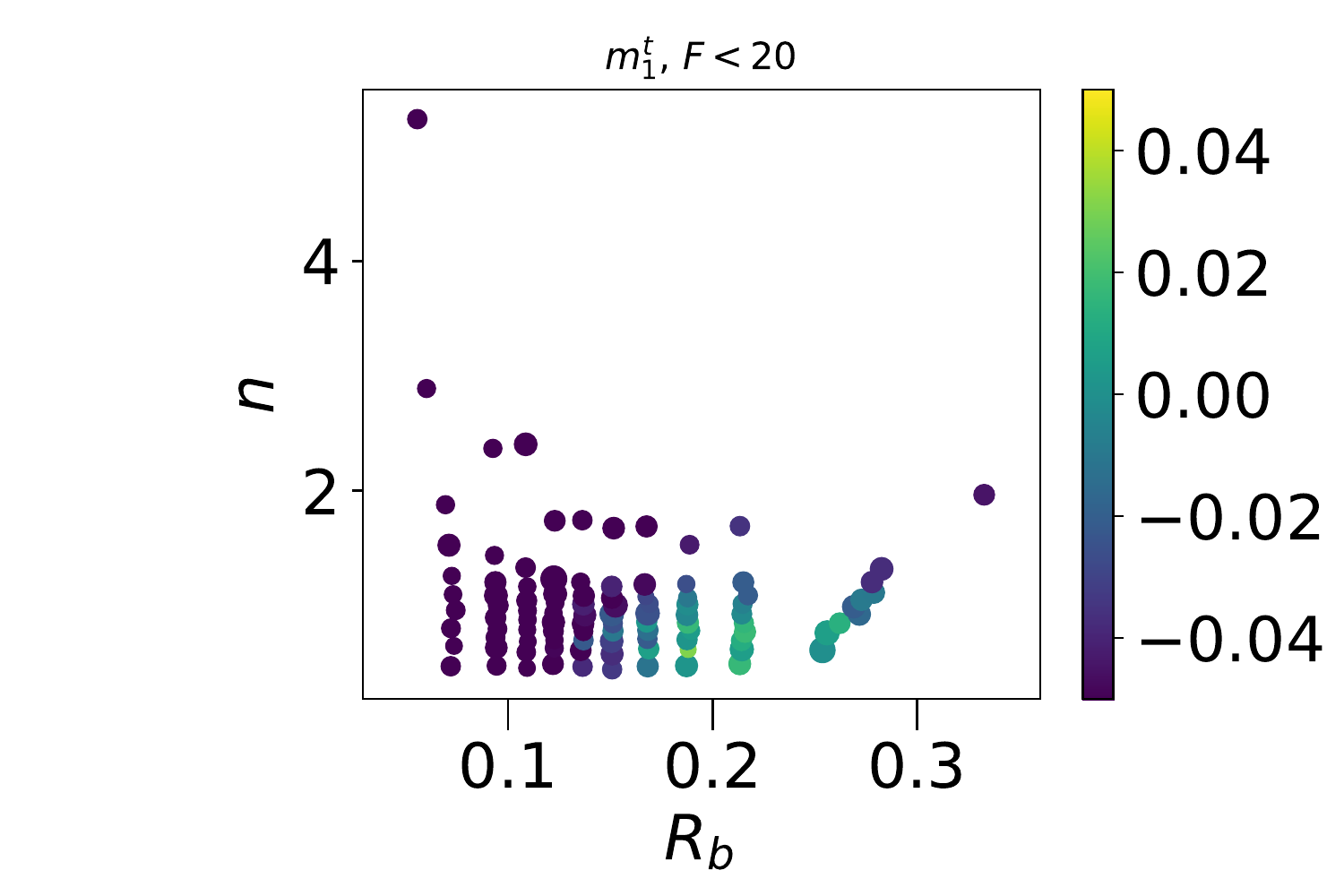}
\includegraphics[width=.95\linewidth]{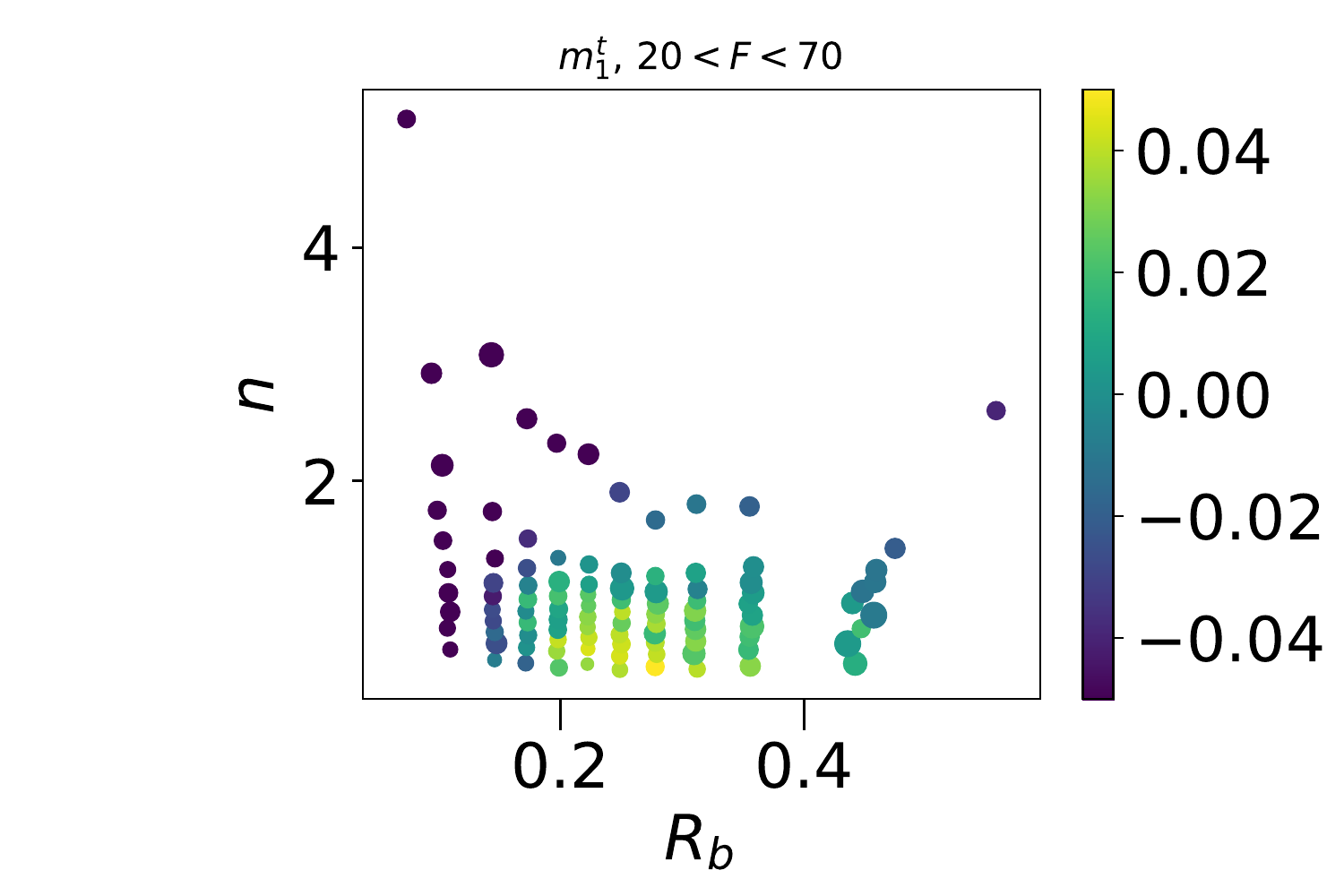}
\includegraphics[width=.95\linewidth]{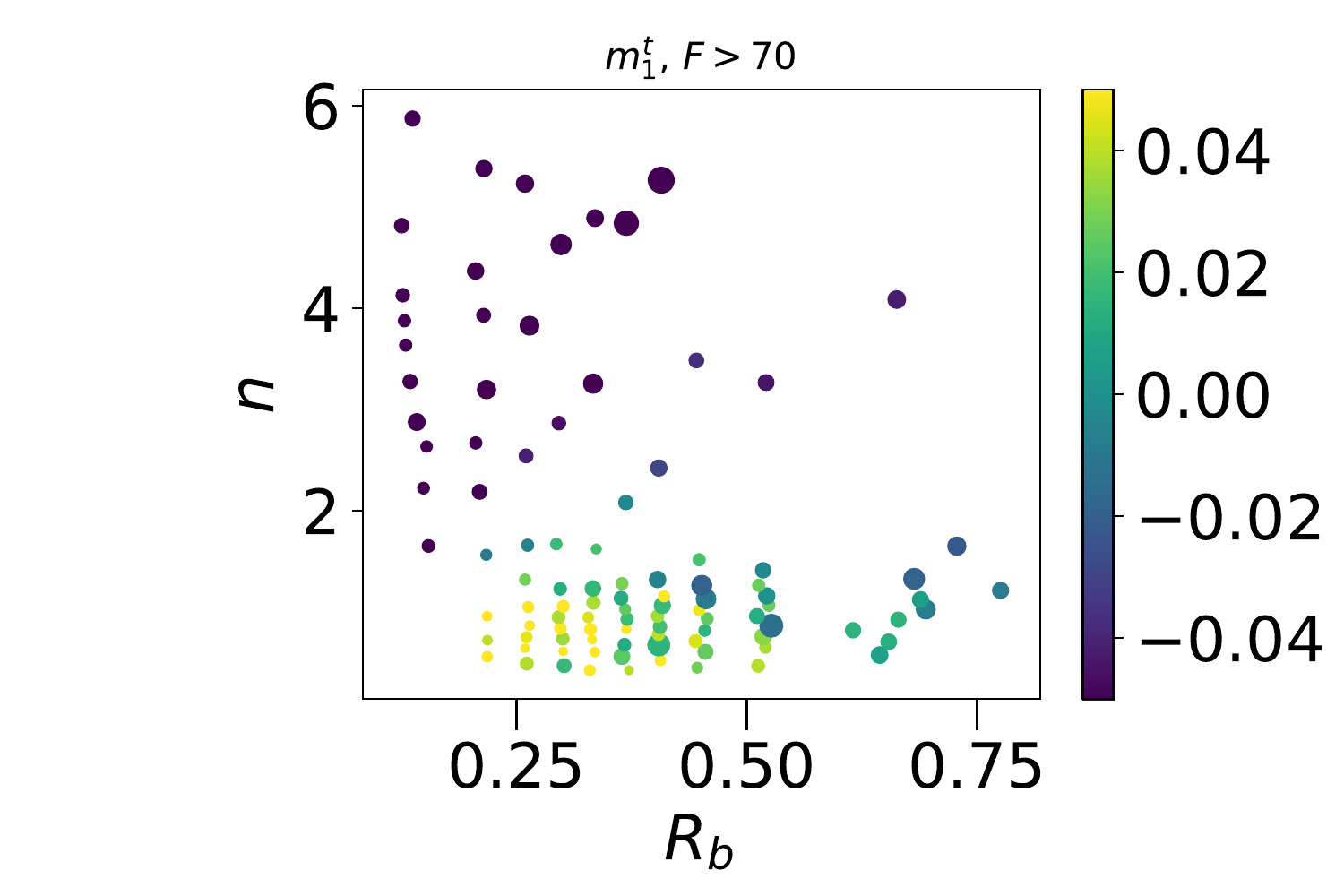}
\caption{Colour-coded true multiplicative shear bias $m_1^{\textrm{t}}$ as a function of input galaxy properties. The $y$-axis shows the Sérsic index $n$, and the $x$-axis is the half-light radius $R_\textrm{b}$. The top, middle, and bottom panels show galaxies with different fluxes, corresponding to $F<20$, $20<F<70,$ and $F>70$, respectively. Each point is to the mean over an equal number of galaxies, and the point size is inversely proportional to the error bar, such that large points are more significant. }
\label{fig:2dm}
\end{figure}

Shear bias depends in a complex way on various properties of the observed galaxy and image properties.
In Paper~I we explore some of these dependences using the same simulated GREAT3 images as in this paper.

In \Fref{fig:2dm} we show an example of shear bias dependences with respect to three galaxy properties. These properties are true parameters from the image simulations, therefore they do not correspond to noisy measurements. Each panel shows that the multiplicative bias $m_1$ depends on the Sérsic index $n$ and the half-light radius $R_\textrm{b}$.
In addition, these dependences change with the galaxy flux $F$ (the three panels show increasing ranges of flux from top to bottom).
We therefore need to know all three quantities to constrain $m_1$, and its
dependence on $n$, $R_\textrm{b}$, and $F$ cannot be separated.  This is just one example,
but it illustrates the
general very complex functional form of shear bias with respect to many galaxy properties. For shear calibration of upcoming high-precision surveys, this sets very high demands on image simulations, which need to densely sample a high-dimensional parameter space of galaxy and image properties.

\section{Deep-learning shear calibration}\label{sec:calibration}

\subsection{Why choose machine learning for shear calibration?}

Sect.~\ref{sec:dependencies} and Paper~I gave a glimpse of shear bias as a very complex, non-linear function acting in a high-dimensional
parameter space of galaxy and image properties. For a successful shear calibration to sub-percent residual biases as is required for
large upcoming surveys \citep[e.g.][]{Massey2013}, this function needs to be modelled very accurately. This is true whether the calibration is performed galaxy by galaxy or globally by forming the mean over an entire galaxy population:
shear bias measured on simulations is always marginalised over some unaccounted-for parameters, explicitly or implicitly. If shear bias depends on some of the unaccounted parameters, then the mean bias is sensitive to the distribution of these parameters over the population used, and a mismatch of this distribution between simulations and observations  produces an incorrect shear bias estimate and calibration. For this reason, it is crucial to model shear bias as a function of a wide range of properties to minimise the effect of the remaining unaccounted-for parameters. With this we would still not have control of the shear bias dependence on the distribution over the remaining unaccounted-for parameters, but we would have already captured the most significant dependences.
Generating a sufficiently large number of image simulations in this high-dimensional, non-separable space obviously sets enormous requirements for computation time and storage for shear calibration.

The exact form of the shear bias function is not interesting per se. In addition, it is very difficult to determine
this function from first principles, based on physical considerations \citep{Refregier2012,Taylor2016,Hall2017,Tessore2019}.
In general, for most simulation-based calibration methods,
empirical functions are fitted. However, we can
make a few very basic and general assumptions about this function. For example, galaxies with similar properties are expected
to have a similar shear bias under a given shape measurement method. Furthermore, this function is expected to be smooth (with a possible stochasticity coming from noise). These basic properties make the shear bias function
ideal to be obtained with machine-learning (ML) techniques.

The problem of estimating the shear bias from  galaxy image parameters requires finely capturing the dependences described above. On the one hand, and from a mathematical point of view, the smoothness of the relationship between shear bias and parameters tells us that shear bias should belong to a smooth low-dimensional manifold. Estimating such a manifold structure then is reduced to some regression problem. On the other hand, the relationship between the measured parameters and the sought-after shear is very intricate, which impedes the use of standard regression methods, but for which ML methods are very well suited.

Machine learning can account for many parameters and model a very complex, high-dimensional
function of shear bias.
The dependence on unaccounted-for parameters of the marginalised shear bias
is expected to be weak,
and the calibration becomes less
sensitive to the particular population that is simulated. If properly designed, we do
not need to know the exact important properties that affect shear bias beforehand because the algorithm can learn the important combination of parameters to constrain
shear bias. We still need to know the property distribution of the observed
galaxies, which are noisy and might be biased. However, as we show below,
the ML training set can be different from the test set to some extent, with
only relatively small calibration bias; see also the discussion in
Sect.~\ref{sec:advantages}.

\subsection{Neural network shear correction}
\label{sec:learning_description}

In this section we describe our new method based on ML that we call neural network shear correction (NNSC). We describe the concepts of the ML approach, the learning, and the calibration steps.
We have made the code publicly available\footnote{https://github.com/arnaupujol/nnsc}.

\subsubsection{Concept}

\begin{figure}
\centering
\includegraphics[width=.95\linewidth]{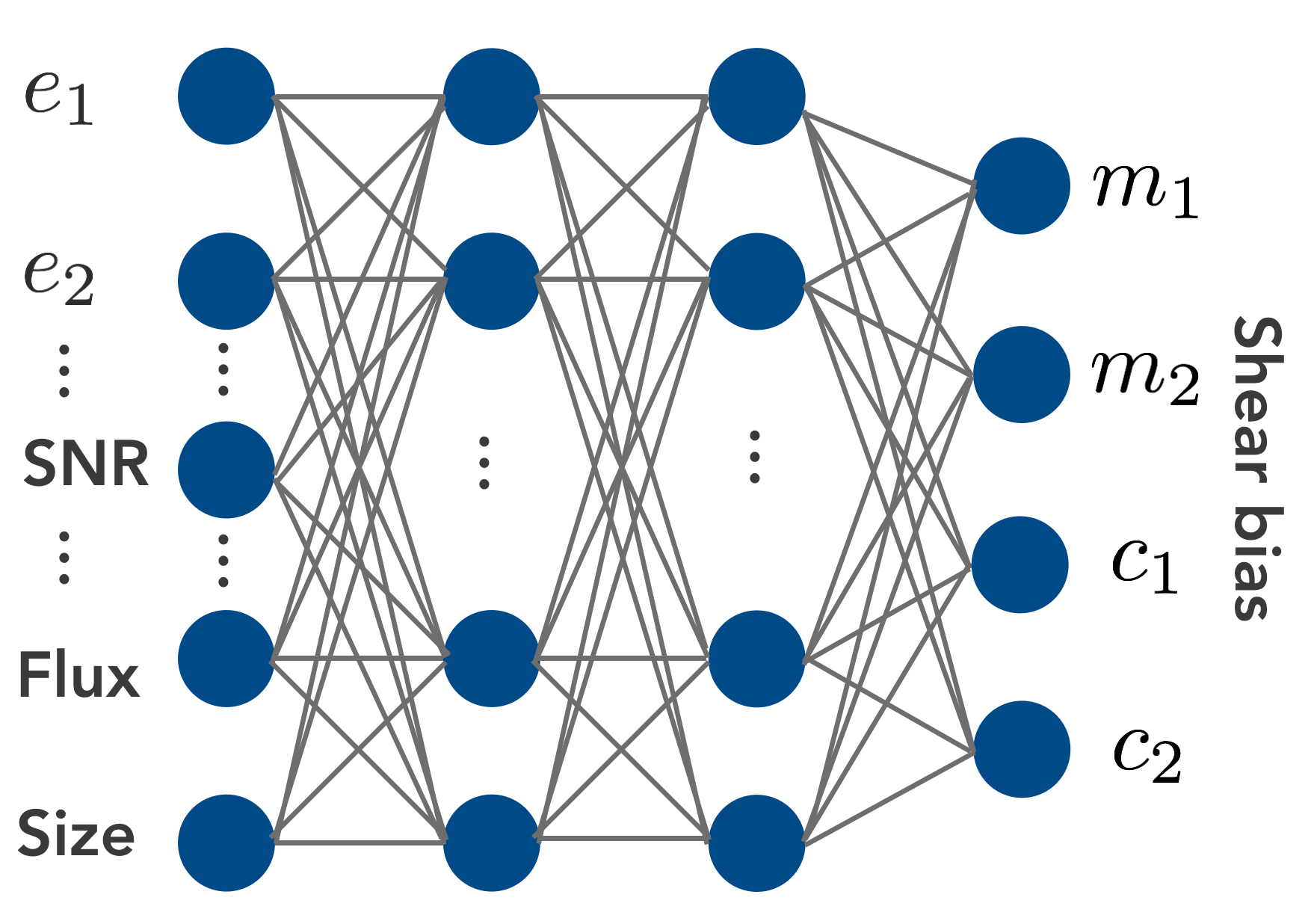}
\caption{Visual schema of the ML approach of the method. A set of measured image properties is used as input features for the neural network. The system is then trained to produce the shear bias parameters as output. }
\label{fig:ml_schema}
\end{figure}

The objective of the ML here is to infer an estimate of the shear bias from a large number of galaxy image measurement parameters. To this end, a deep neural network (DNN), and more precisely, a multi-layer feed-forward neural network, is particularly well adapted to solve the underlying regression problem.

More precisely, a feed-forward neural network is composed of $L$ layers, taking measured image properties as inputs and providing the shear bias as output. The resulting network aims at mapping the relationship between the measured properties of the galaxy images and the shear bias. The parameters of the network are then learnt in a supervised manner by minimising the residual between the true and the estimated shear bias. The shear bias is estimated individually for each galaxy given their measured properties.

If $x_i$ denotes a vector containing $m$ measured properties $x_i[j]$ for $j=1,\cdots,m$  for a single galaxy $i$, then the output of the first layer $\ell = 1$ of the neural network is defined by some vector $h_j^{(1)}$ of size $m_1$,
\begin{equation}
h_i^{(1)} = \mathcal{A}\left({\bf W}^{(1)}x_i + b^{(1)}\right),
\end{equation}
where ${\bf W}^{(1)}$ ($b^{(1)}$) stands for the weight matrix (the bias vector) at layer $\ell=1$, $x_i$ is the vector of the galaxy with index $i,$ and $j$ is the index referring to a measured property.
The term $\mathcal{A}$ is the so-called activation function, which applies entry-wise on its argument. For a layer $\ell = n$, the output vector of size $m_n$ is defined as
\begin{equation}
h_i^{(n)} = \mathcal{A}\left({\bf W}^{(n)}h_i^{(n - 1)} + b^{(n)}\right).
\end{equation}
For $\ell = 5$, the output vector $h_j^{(L)}$ stands for the estimated shear bias components. Only in this last layer is the activation function not used, so that we gain a linear function to obtain the shear bias components. A visual schema of the method is shown in Fig.~\ref{fig:ml_schema}.

\subsubsection{Learning}

The learning stage amounts to estimating the parameters $\left\{ {\bf W}^{(\ell)}, b^{(\ell)}  \right \}_{\ell = 1,\cdots,L}$ by minimising the following cost function defined as some distance (the $\chi^2$) between the shear bias components and their estimates:

\begin{equation}
C = \frac{1}{b_\textrm{s}} \sum_{i=0}^{b_\textrm{s}} \sum_{\alpha = 1}^2 (m_{i,\alpha}^\textrm{t} - m_{i,\alpha}^\textrm{e})^2 + (c_{i,\alpha}^\textrm{t} - c_{i,\alpha}^\textrm{e})^2,
\label{eq:cost_function}
\end{equation}
where $m_\alpha^\textrm{t}$, $c_\alpha^\textrm{t}$ and $m_\alpha^\textrm{e}$, $c_\alpha^\textrm{e}$ are the true and estimated $\alpha$th component of the shear multiplicative and additive bias, respectively, and $b_\textrm{s}$ is the number of objects used in each training step (also referred to as the batch size).
We use as true shear bias the values from equations (\ref{eq:shear_response}) and (\ref{eq:ind_add_bias}) obtained as described in Sect.~\ref{sec:shear_bias_measurement} and presented in \cite{Pujol2018}. The NNSC learns to estimate these shear biases.

We used $m = 27$ measured properties used as input for the model as described in Sect.~\ref{sec:input_data}, and details of the network architecture and the implementation of the learning stage are given in Appendix~\ref{sec:training}.

\subsubsection{Calibration}

The NNSC method estimates the shear responses and biases of individual galaxies from the measurements of $27$ properties applied to the images. The shear bias and the corresponding calibration were made for a previously chosen shape measurement. Any shape measurement algorithm can be chosen for this purpose. We used the estimation from the \ksb\ method using the software \shapelens. When these estimations were completed, we applied the shear calibration over the statistics of interest, in our case, the estimated shear from Eq.~\ref{eq:g_relation}.
The bias calibration was applied as $\langle \mat R\rangle^{-1} \langle \bm e^{\rm obs} - \langle \bm c \rangle \rangle $, where $\mat R$ and $\bm c$ are the estimated average shear response and additive bias, respectively \citep[see][]{Sheldon2017}. This calibration is similar to other common approaches, and we expect similar behaviours for the post-calibration bias as discussed in \cite{Gillis2019}.

Our method gives estimates for the individual shear bias of objects, in common with the new method \metacal. However, the two methods are very different. While NNSC relies on image simulations for a supervised ML approach, \metacal\ uses the data images themselves to obtain the individual shear responses. To do this, the original data images are deconvolved with an estimated PSF, and after some shear is applied, they are re-convolved  with a slightly higher PSF. Because this method is very complementary with respect to NNSC and has recently been used in surveys such as the Dark Energy Survey (DES) \citep{Zuntz2017}, we used it in this study for a calibration comparison of both models. For more details of \metacal,\ we refer to a description of the implementation in Appendix \ref{sec:metacal} and the original papers \citep{Huff2017,Sheldon2017}.

\begin{table*}
    \centering
    \begin{tabular}{|c|c|c|}
    \hline
        \gfit & \pse & \ksb \\
        \hline
        Galaxy ellipticity $e_{1,\gfit}$, $e_{2,\gfit}$ & Galaxy flux $F_{\rm{out}}$ & Ellipticity $e_{1,\ksb}$, $e_{2,\ksb}$ \\
        Axis ratio  & Galaxy size & Ellipticity with respect to the PSF $e_{+,\ksb}$, $e_{\times,\ksb}$ \\
        Orientation angle & $\rm{S/N}_{\rm{obs}}$ & Axis ratio $q_\ksb$ \\
        Galaxy flux & Galaxy magnitude & Orientation angle $\beta_\ksb$ \\
        Disc radius & PSF flux & Window function size \\
        Bulge radius & PSF size & S/N \\
        Disc fraction & PSF S/N & \\
        Number of $\chi^2$ evaluations & PSF magnitude & \\
        Noise level & PSF FWHM & \\
    \hline
    \end{tabular}
    \caption{Measured properties for the training process of NNSC.}
    \label{tab:props}
\end{table*}

\section{Data}\label{sec:data}

\subsection{Image simulations}

We  considered two sets of \galsim\ simulations \citep{Rowe2015}. They
correspond essentially to the control-space-constant (CSC) and
real-space-constant (RSC) branch simulated in  GREAT3 \citep{Mandelbaum2014},
with some modifications to ensure precise measurement of the shear response as
prescribed in PKSB19.

The CSC branch contains galaxies with parametric profiles (either a single Sérsic or a de Vaucouleurs bulge profile to which an exponential disc was added) obtained from fits to Hubble Space Telescope (HST) data from the COSMOS survey with realistic selection criteria \citep{Mandelbaum2014}. This data set is intended to provide a realistic distribution of galaxy properties (in particular in terms of morphology, size, ans S/N), which we therefore used for training and testing our calibration network.

As for GREAT3, the two million galaxy images were divided into 200 images  of 10000 galaxies, to each of which a different pre-defined shear and PSF was applied. Each galaxy was randomly oriented, and its orthogonal version was also included in the data set to allow for nulling the average intrinsic ellipticity. Out of these 200 images, we selected a first set for training and a second set for testing and comparing calibration approaches. For the training set, we followed the approach of PKSB19 described above to obtain an estimate of the true shear response that needed to be learnt. For each  galaxy in the training set, five sheared versions were simulated keeping PSF and noise realisations the same. The  shear $\bm g$ for each galaxy was chosen as $\bm g_i = \{(g_{1}, g_{2})_i\} = \{ (0, 0), (\pm 0.02, 0), (0, \pm 0.02) \} $.

To further investigate the effect of model bias on our procedure, the network predictions were also tested for more realistic galaxies simulated as in the RSC branch of GREAT3. These galaxies are based on actual observations from the HST COSMOS  survey,  fully deconvolved with the HST PSF (see the procedure in \citet{Mandelbaum2013}), before we applied random rotation, translation, and the prescribed shear followed by convolution with the target PSF and  resampling in the target grid. In this scenario, the same procedure as for CSC was followed to obtain estimates of the shear response for these realistic galaxies, which were then compared to the network predictions based on the CSC training set.

\subsection{Learning input data}\label{sec:input_data}

The image properties that are used to estimate the shear bias can be chosen depending on the interest. The NNSC
learns to estimate shear bias as a function of these properties, which means that the more properties we use,
the more capable the NNSC is to learn complex dependences (if we use the appropriate training).

We used $27$ measured properties as input for the NNSC. These properties correspond
to the output from the \gfit\ \citep{Gentile2012, Mandelbaum2015} software (properties such as ellipticities, fluxes, sizes, fitted disc fraction, and other fitting statistics), the \shapelens\ \citep{Viola2011} \ksb\ implementation
 (ellipticities, S/N, and the size of the window function) and \pse\  \citep{Bertin1996} software (properties such as flux, size, S/N, and magnitude from both the galaxy and the PSF).
We refer to Paper I for the details on the algorithms and implementations and to Table~\ref{tab:props} for the list of measured properties we used for the training.
In the following we report the results obtained with the selected network as described in Appendix~\ref{sec:training} associated with the superscript "fid", referring to the fiducial implementation of the method.

\section{Results}\label{sec:results}

\subsection{Bias predictions}\label{sec:bias_preds}

\begin{figure}
\centering
\includegraphics[width=.98\linewidth]{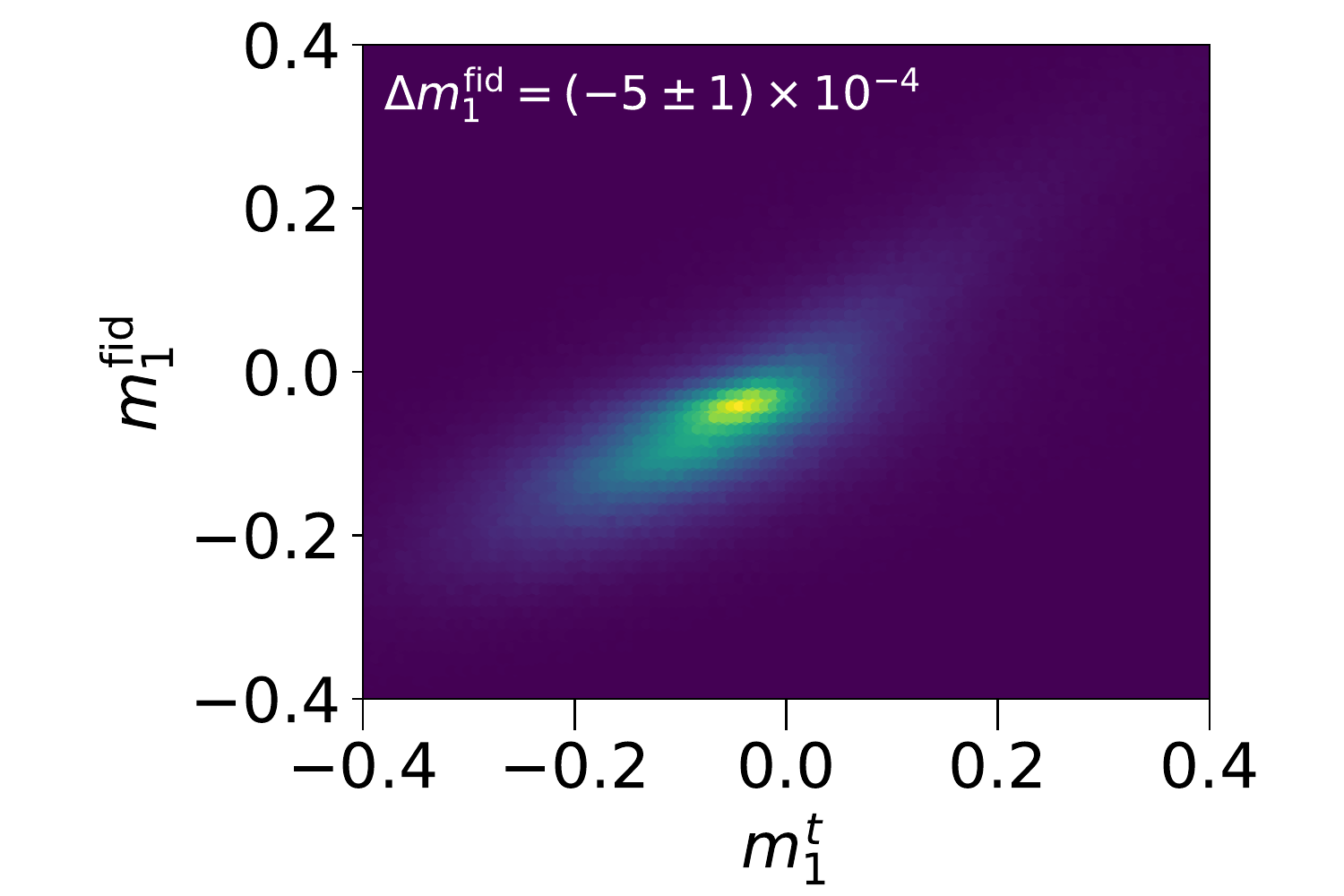}
\includegraphics[width=.98\linewidth]{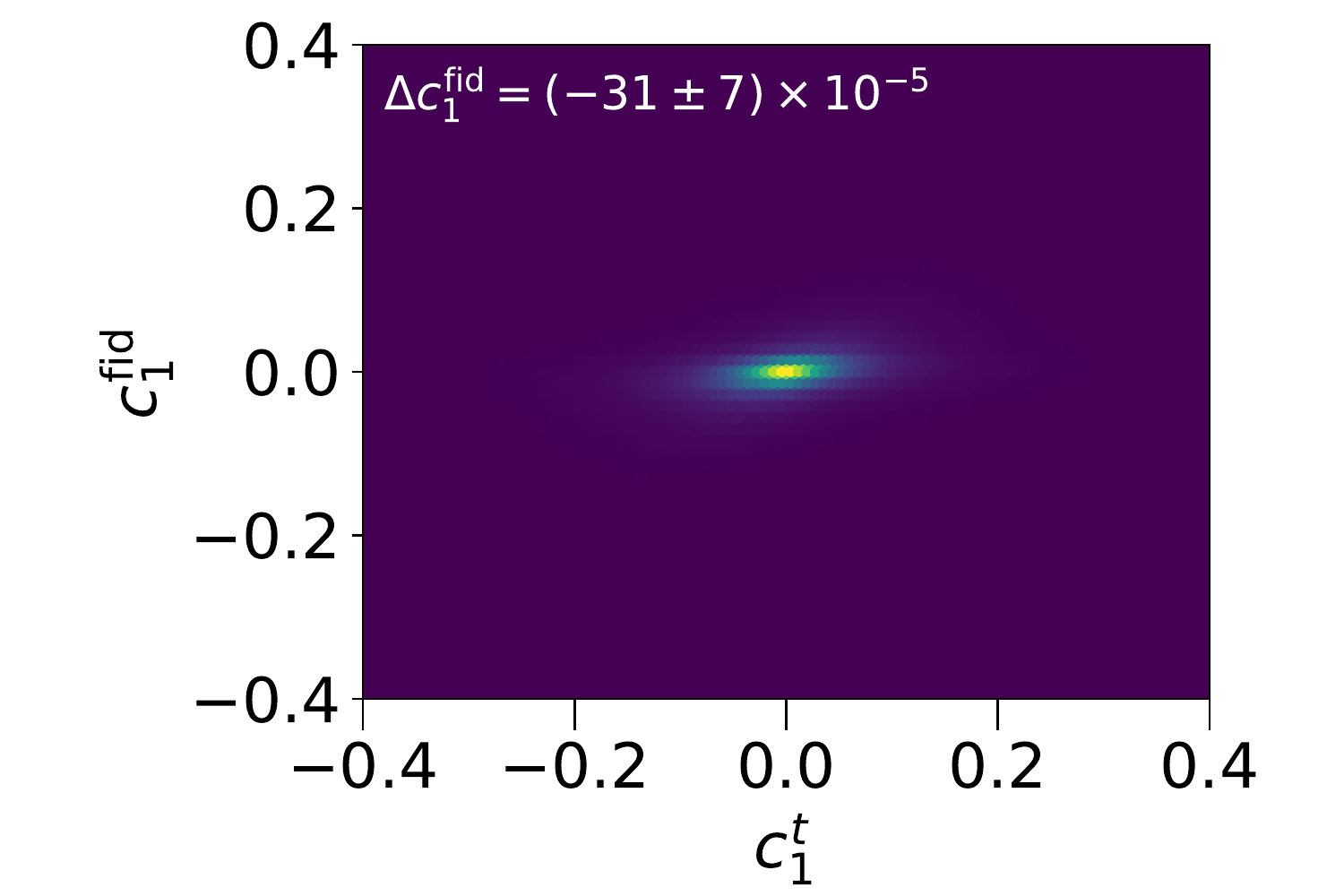}
\caption{Comparison between true and estimated shear bias. The multiplicative shear bias $m_1$ is shown in the top panel, and in the bottom panel, we show the additive bias $c_1$.
}
\label{fig:mest_mtrue}
\end{figure}

\begin{figure*}
\centering
\includegraphics[width=.48\linewidth]{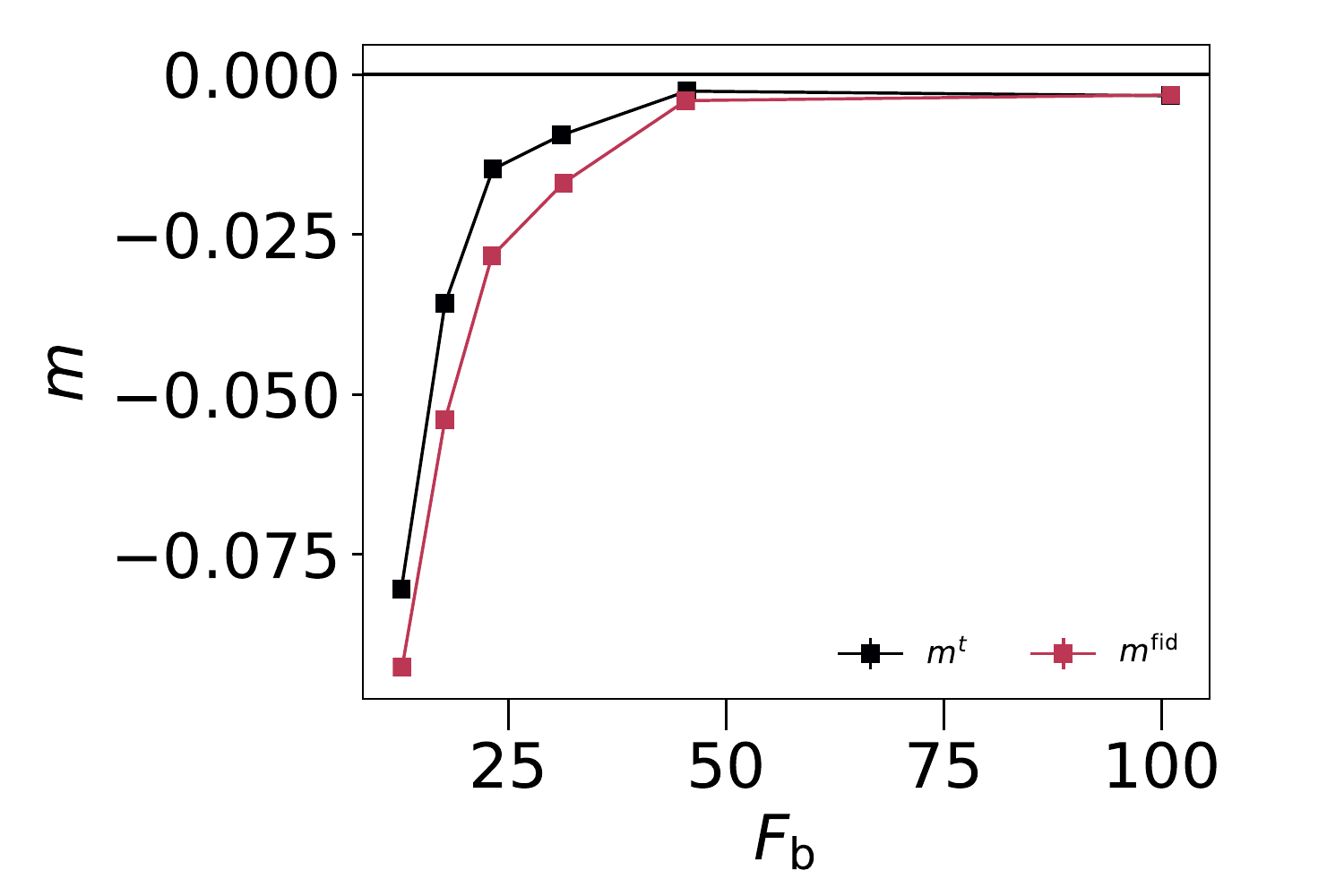}
\includegraphics[width=.48\linewidth]{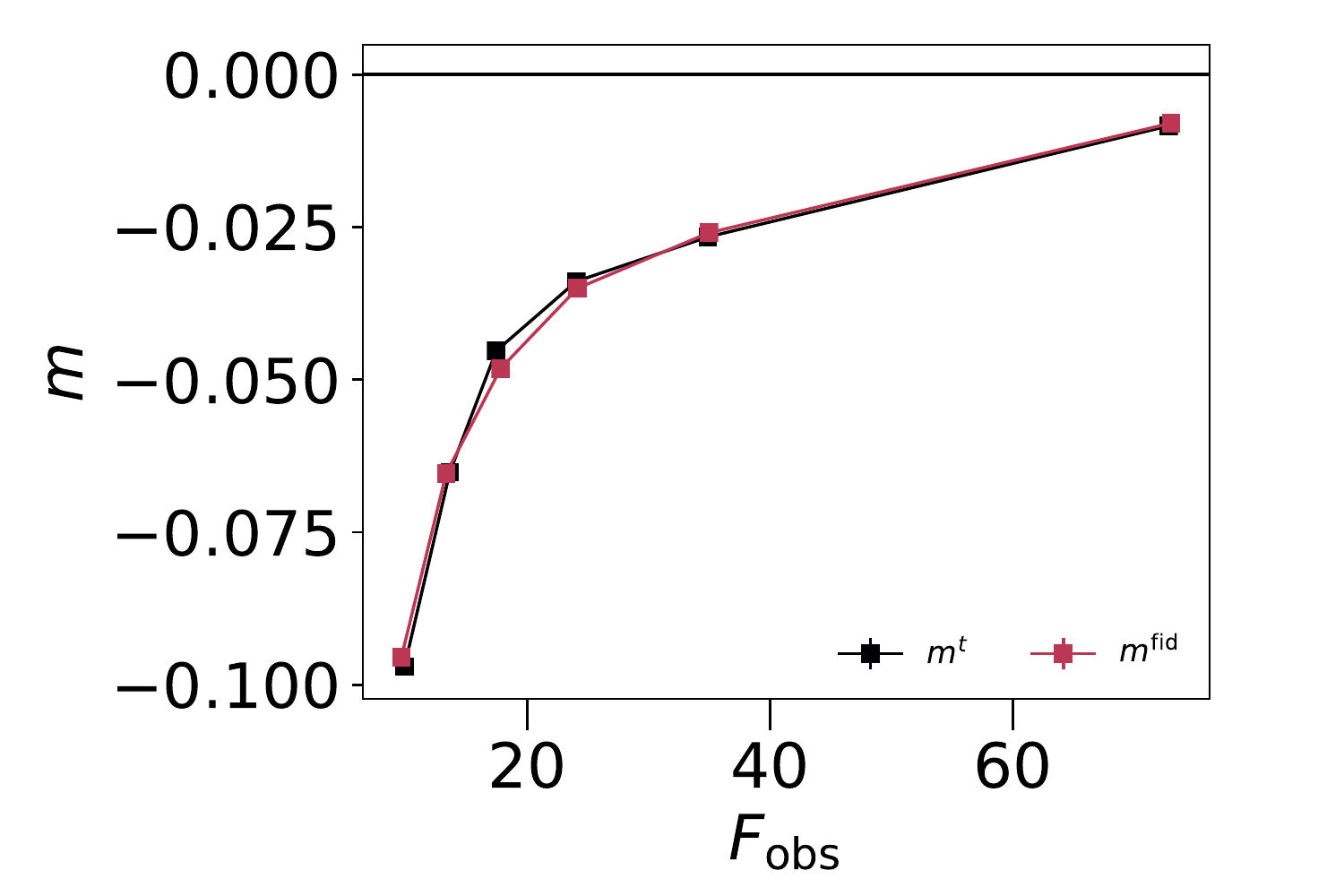}
\includegraphics[width=.48\linewidth]{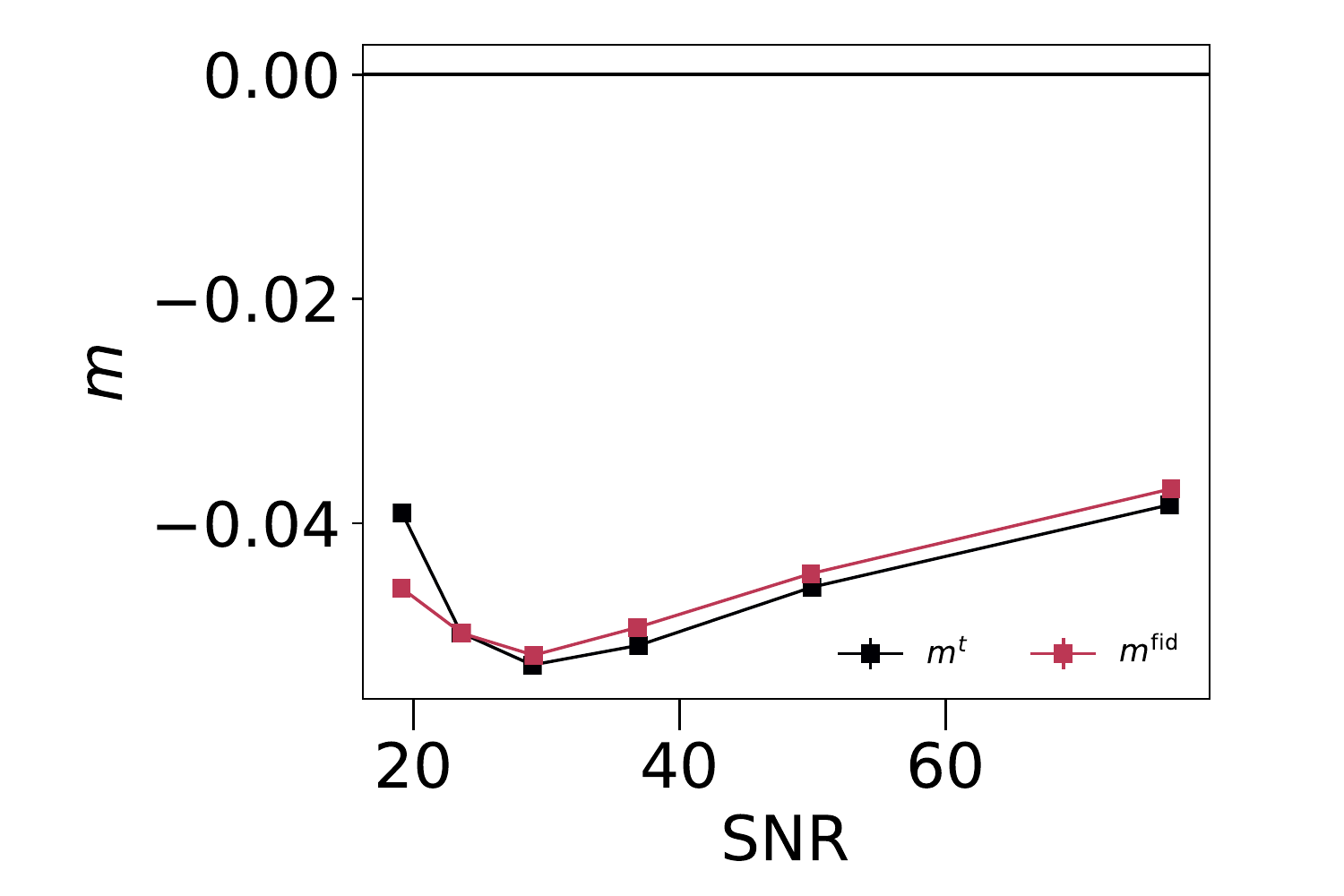}
\includegraphics[width=.48\linewidth]{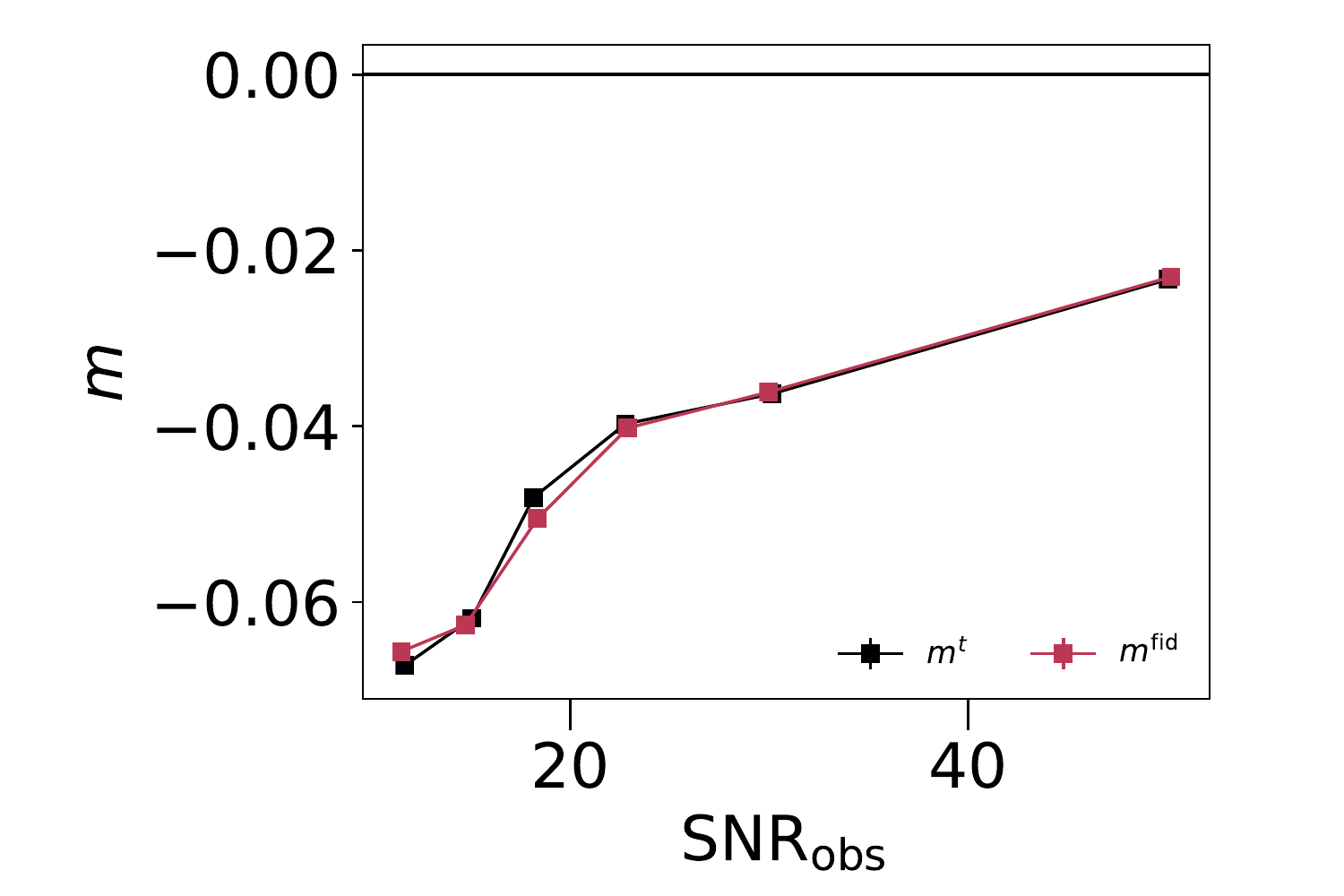}
\caption{Comparison of estimated vs. true multiplicative shear bias $m_1$ as a function of several properties. The top panels show $m$ as a function of galaxy flux, and the bottom panels show $m$ as a function of S/N. The left panels show input simulation properties from the galsim parameters, and the right panels show the measured parameters from \pse\  that were used as inputs in training the NN. }
\label{fig:1d_pred_m_comp}
\end{figure*}

In Fig. \ref{fig:mest_mtrue} we show the distribution of estimated and true shear biases in the validation set of the CSC branch. We show $m_1$ (top panel) and $c_1$ (bottom panel), but similar results are found for $m_2$ and $c_2$. The estimated and true biases are correlated, although the relation is scattered. The value distribution is also narrower for the estimated than for the true biases because the estimated bias is a function of the measured parameters with no noise stochasticity. This has been learned from the stochastic true values that are affected by noise (which is the main cause of the scatter of the true-bias values), but the estimated function is not stochastic.

The errors on the estimated average biases, defined as $\Delta m_{1,2} = \langle m_{1,2}^{\rm{fid}} - m_{1,2}^{\rm{t}} \rangle$ (and analogous for $c_{1,2}$),  are $\Delta m_1 = (4.9 \pm  1.1) \times 10^{-4}$ and $\Delta c_1 = (-3.1 \pm 0.7) \times 10^{-4}$ (similar values are found for the second components, with $\Delta m_2 = (0.0 \pm  1.1) \times 10^{-4}$ and $\Delta c_2 = (1.6 \pm 0.7) \times 10^{-4}$).
These values are well below the Euclid requirements ($\Delta m < 2 \times 10^{-3}$ and $\Delta c < 5 \times 10^{-4}$), although a proper test of Euclid simulations should be made to quantify the performance for this mission and to quantify the effct on post-calibration bias, for which the requirements are set \citep{Massey2013}. However, this performance was obtained using only $128,000$ objects with $\sim 15$ CPU training hours.

To obtain this precision, we used a validation set of about $1,800,000$ objects. According to the results from PKSB19, we expect an error on the mean bias of $\sim 3 \times 10^{-4}$. However, here we show the error on $m_1^t - m_1^{\rm est}$. If these two quantities are correlated (as they are), the error on their difference can be smaller, as we show.

\subsection{1D dependences}

\begin{figure*}
\centering
\includegraphics[width=.48\linewidth]{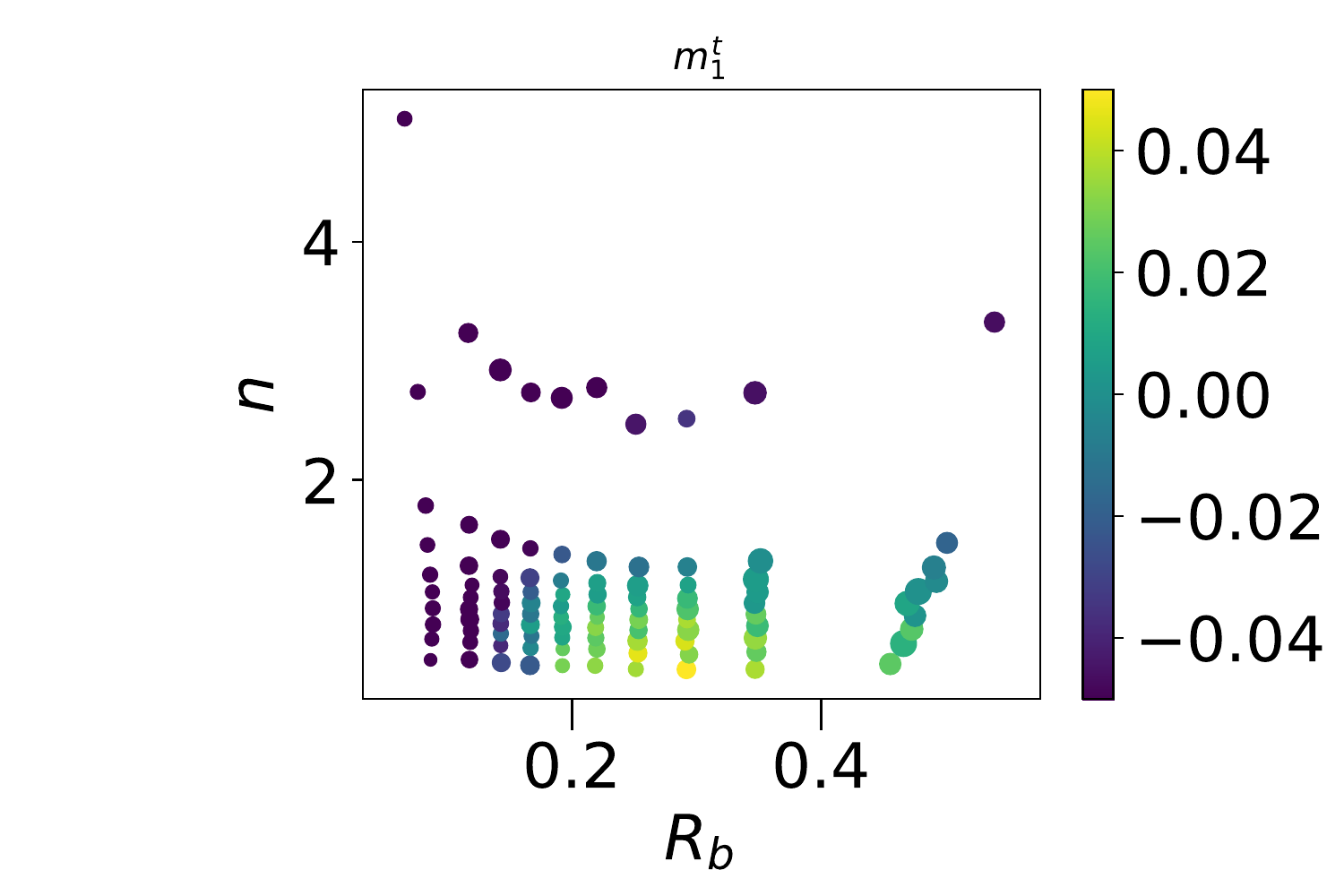}
\includegraphics[width=.48\linewidth]{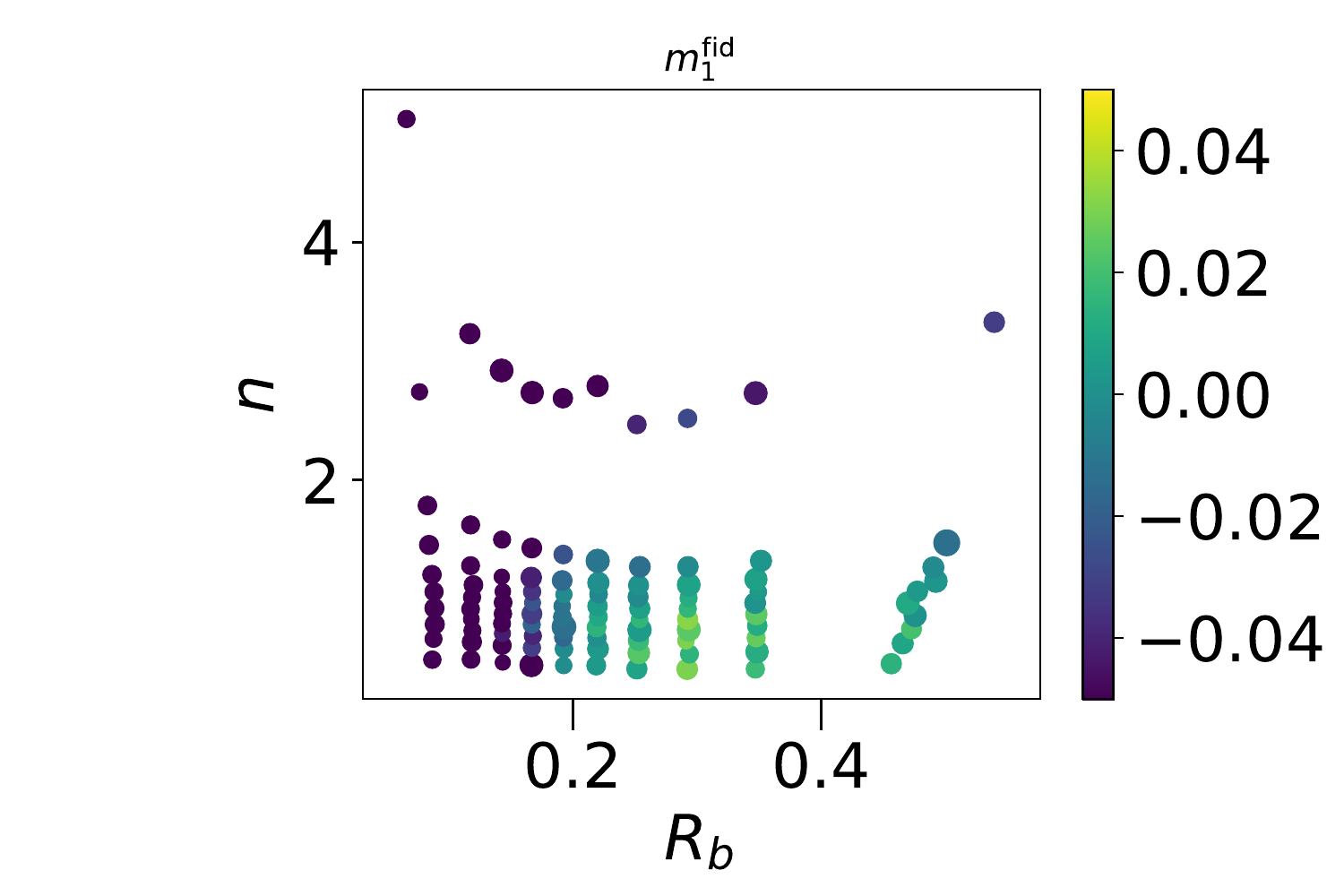}
\includegraphics[width=.48\linewidth]{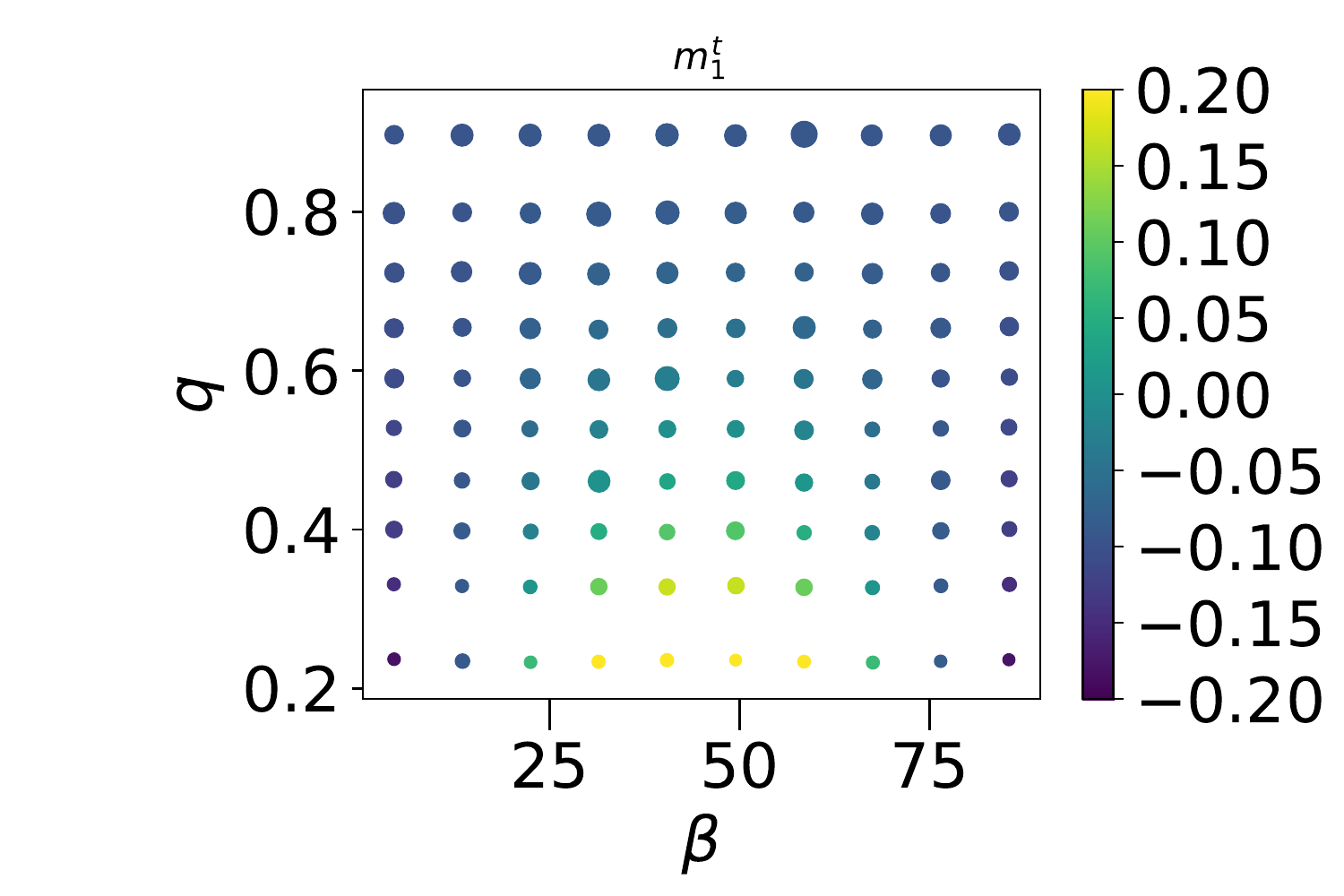}
\includegraphics[width=.48\linewidth]{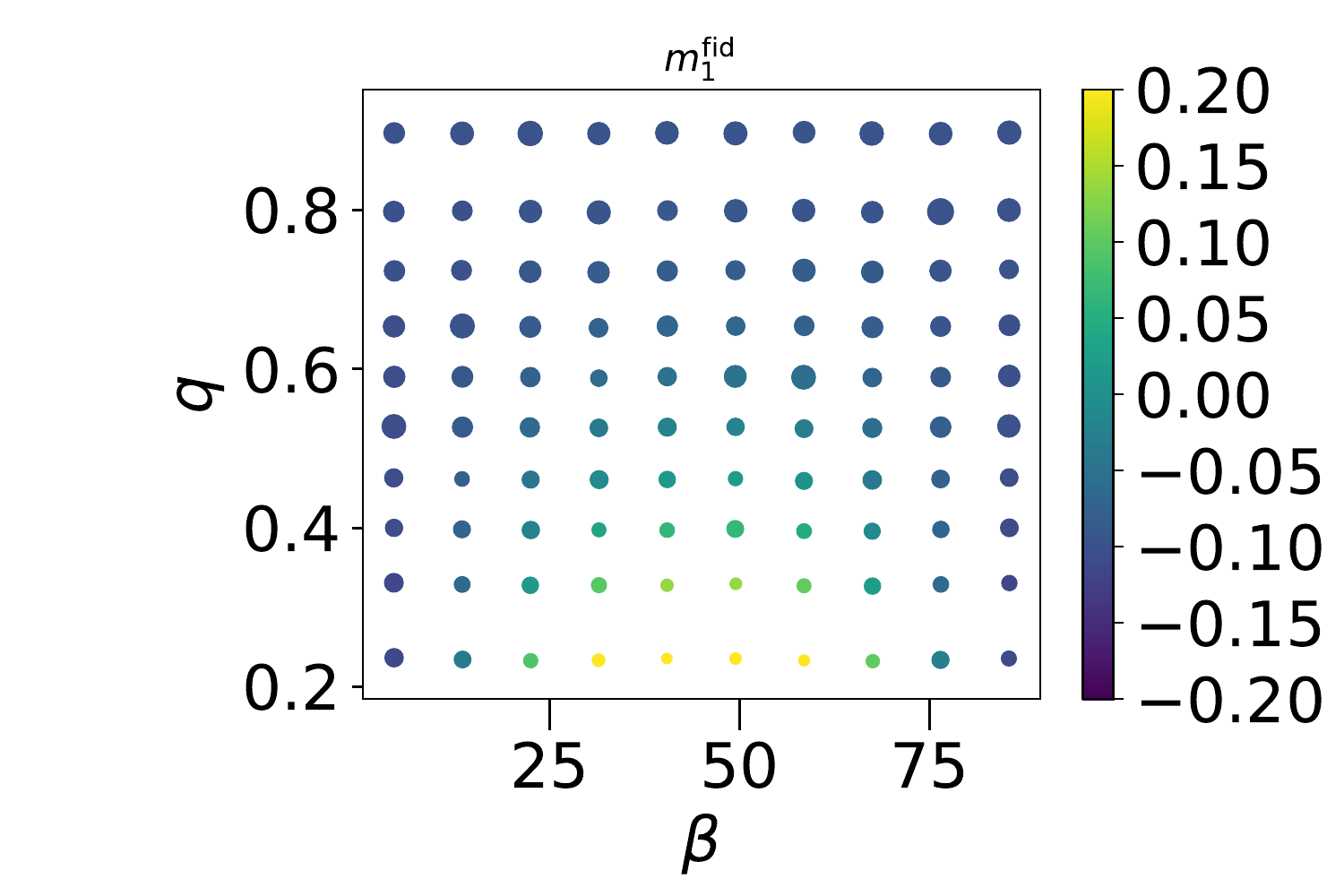}
\includegraphics[width=.48\linewidth]{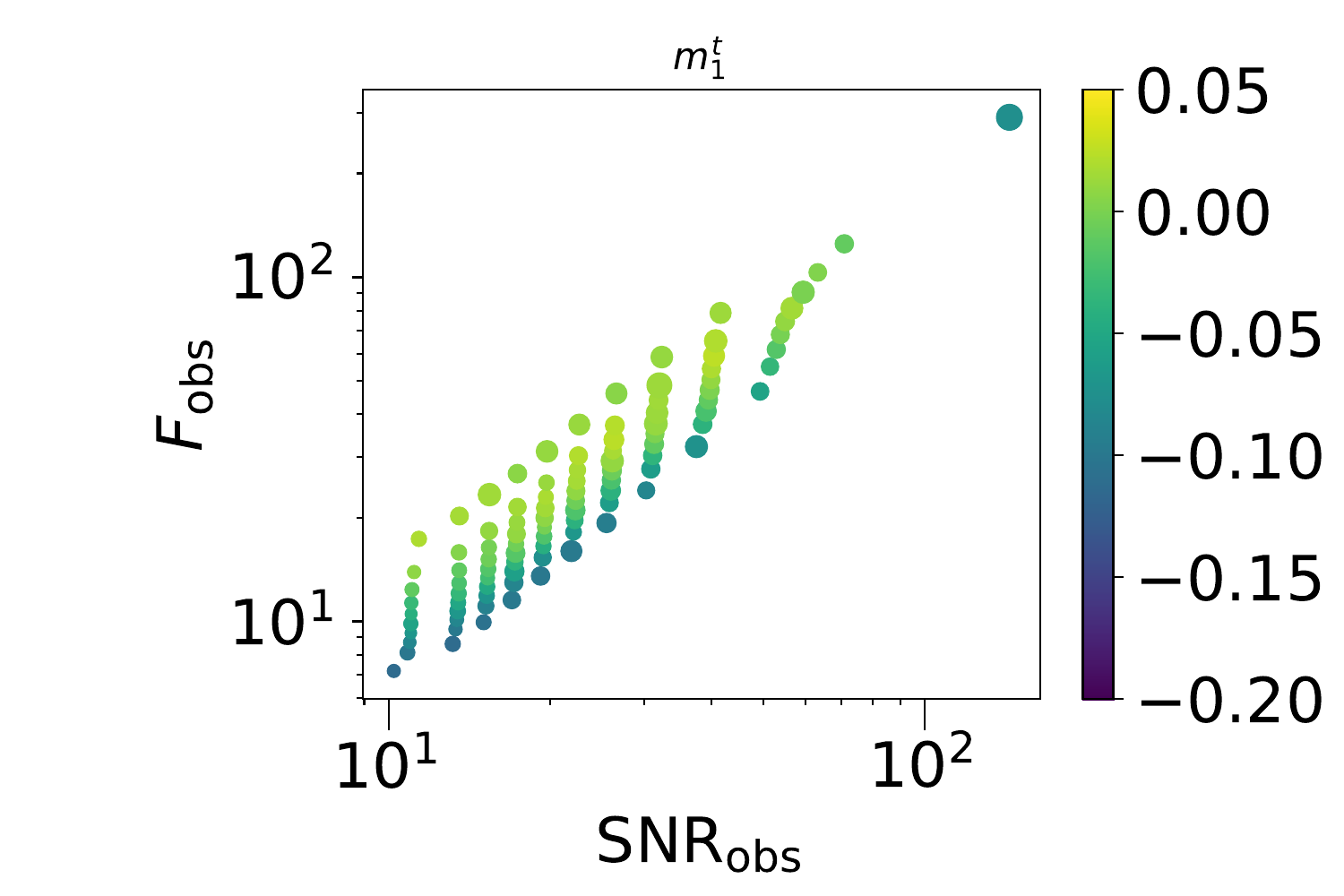}
\includegraphics[width=.48\linewidth]{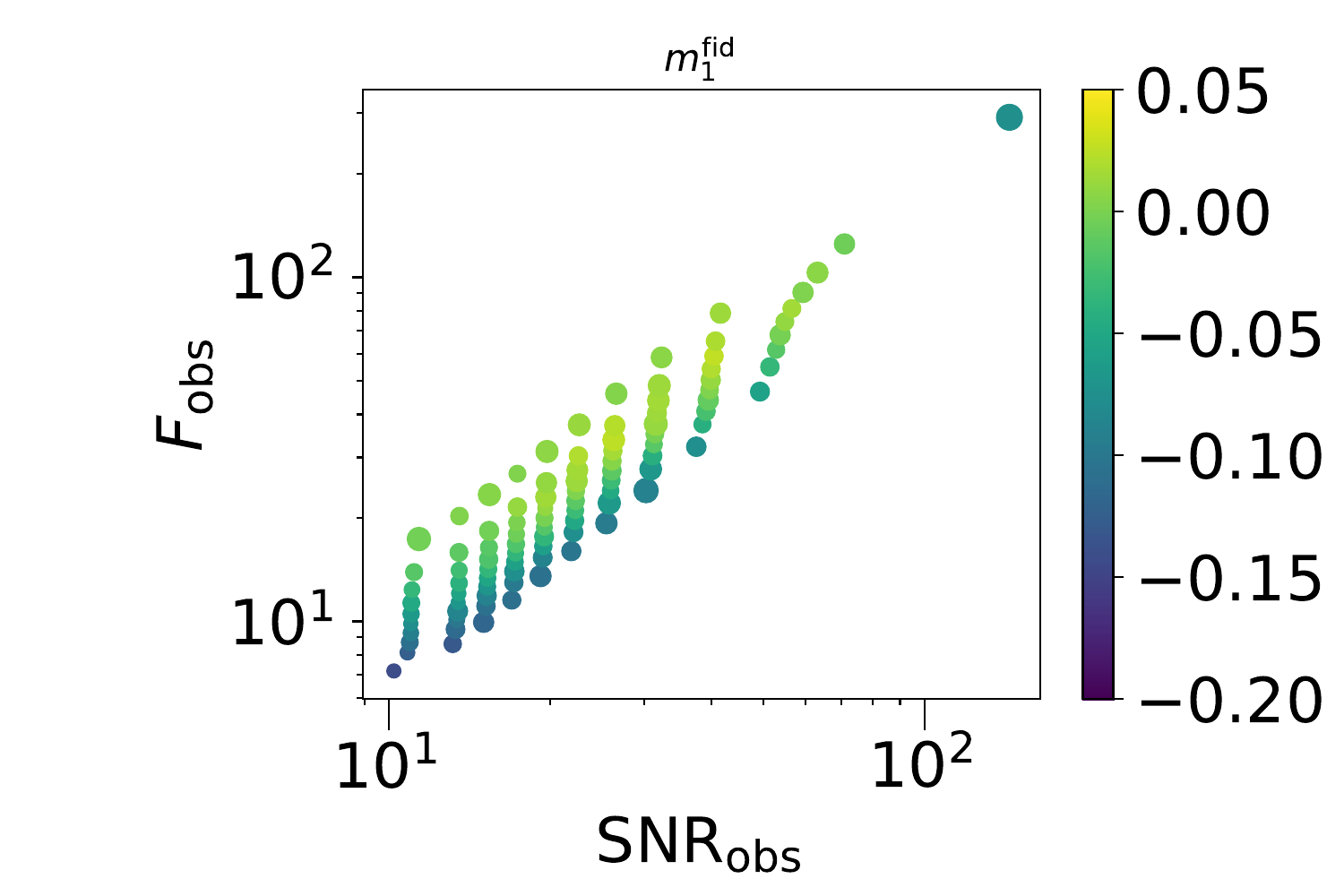}

\caption{Simultaneous comparison of true (left panels) and estimated (right panels) multiplicative shear bias $m_1$ as a function of two properties. The multiplicative shear bias is represented in colours. Each point is to the mean over an equal number of galaxies, and the point size is inversely proportional to the error bar, so that large points are more significant. Top panels: Dependences on Sérsic index $n$ and bulge half-light radius in the simulation. Middle panels: Dependences on intrinsic ellipticity modulus $q$ and orientation angle $\beta$ in the simulation. Bottom panels: Dependences on the measured flux and S/N from \pse.
}
\label{fig:2d_m_comp}
\end{figure*}

In Fig. \ref{fig:1d_pred_m_comp}  we show some examples of shear bias dependences for different cases. In black we show the true multiplicative bias obtained as described in Sect. \ref{sec:shear_measurements} that was used for the supervised training.
The dark red line corresponds to the performance of the NNSC estimation, referred to as $m^{\rm{fid}}$ because it represents the fiducial training parametrisation used for this paper (for other parametrisations, see Appendix \ref{sec:training}).
In the left panels we show dependences on input simulation parameters. These are parameters that were used to generate the image simulations with \galsim, but they were not used for NNSC. The training therefore does not have access to these properties. In the right panels we show dependences on measured parameters that were used for the training. The top panels show the $m$ dependence on galaxy flux, and the bottom panels the dependence on S/N. The excellent performance in the right panels shows that the training correctly reproduces the dependences on the measured parameters that were used as the training input. The left panels show that although the performance is not perfect, the measured parameters used in NNSC capture enough information to reproduce the dependences with good precision\footnote{Only single Sérsic galaxies were used in the top left panel because the true bulge flux only contains a fraction of the flux information for disc galaxies. For this reason, the average difference between estimated and true bias is different than in the rest of the panels, where the whole population was used. }.

\begin{figure*}
\centering
\includegraphics[width=.48\linewidth]{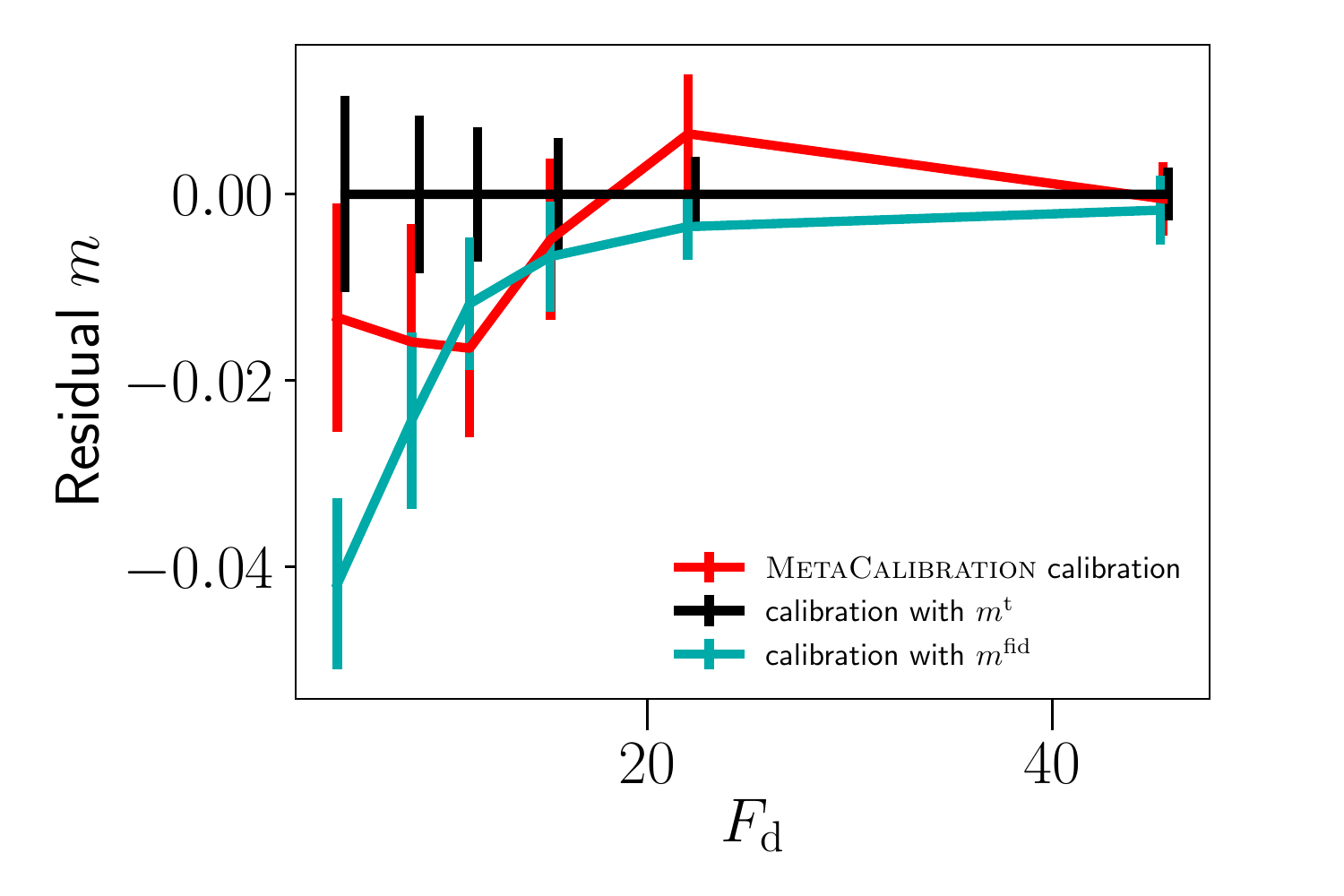}
\includegraphics[width=.48\linewidth]{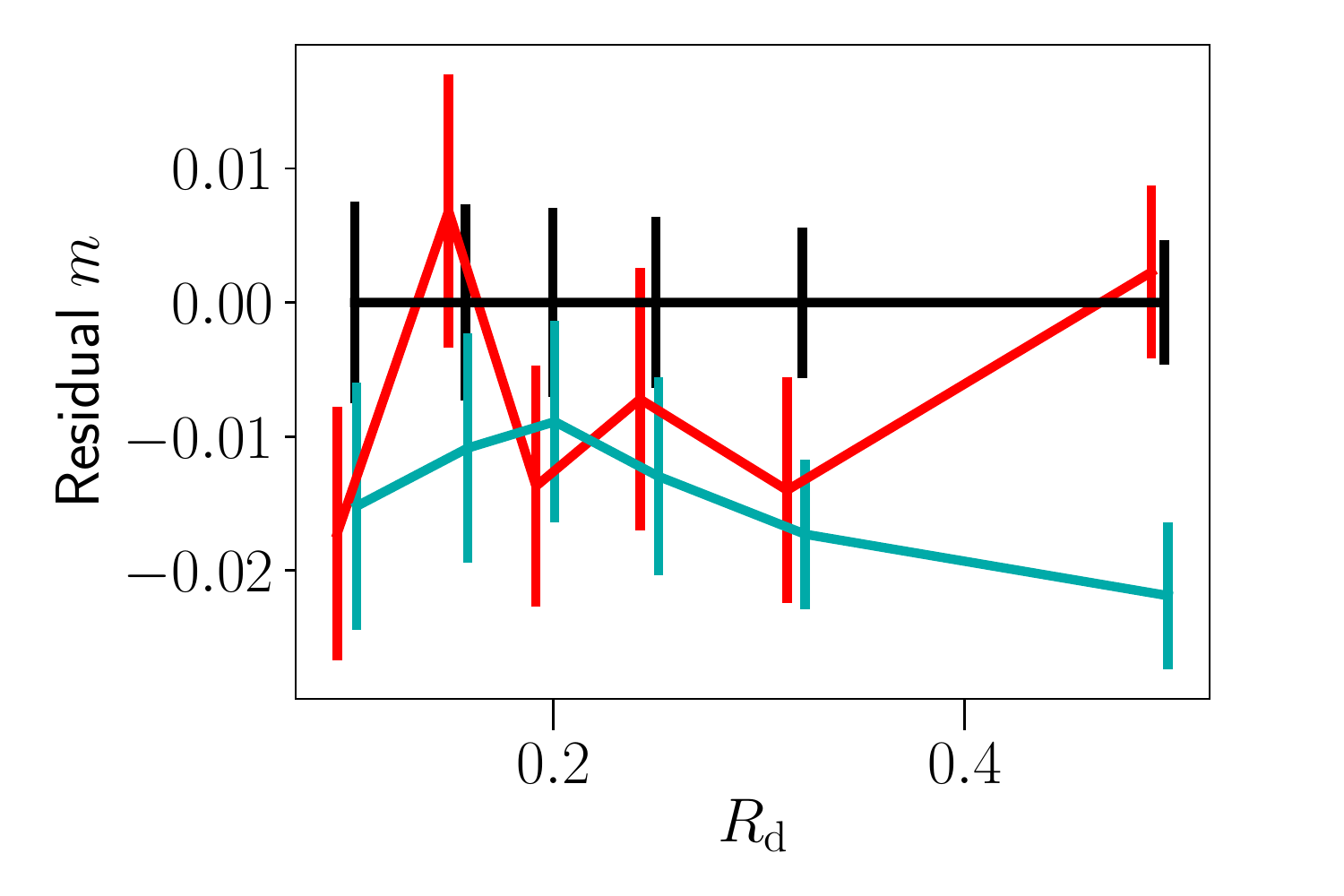}
\includegraphics[width=.48\linewidth]{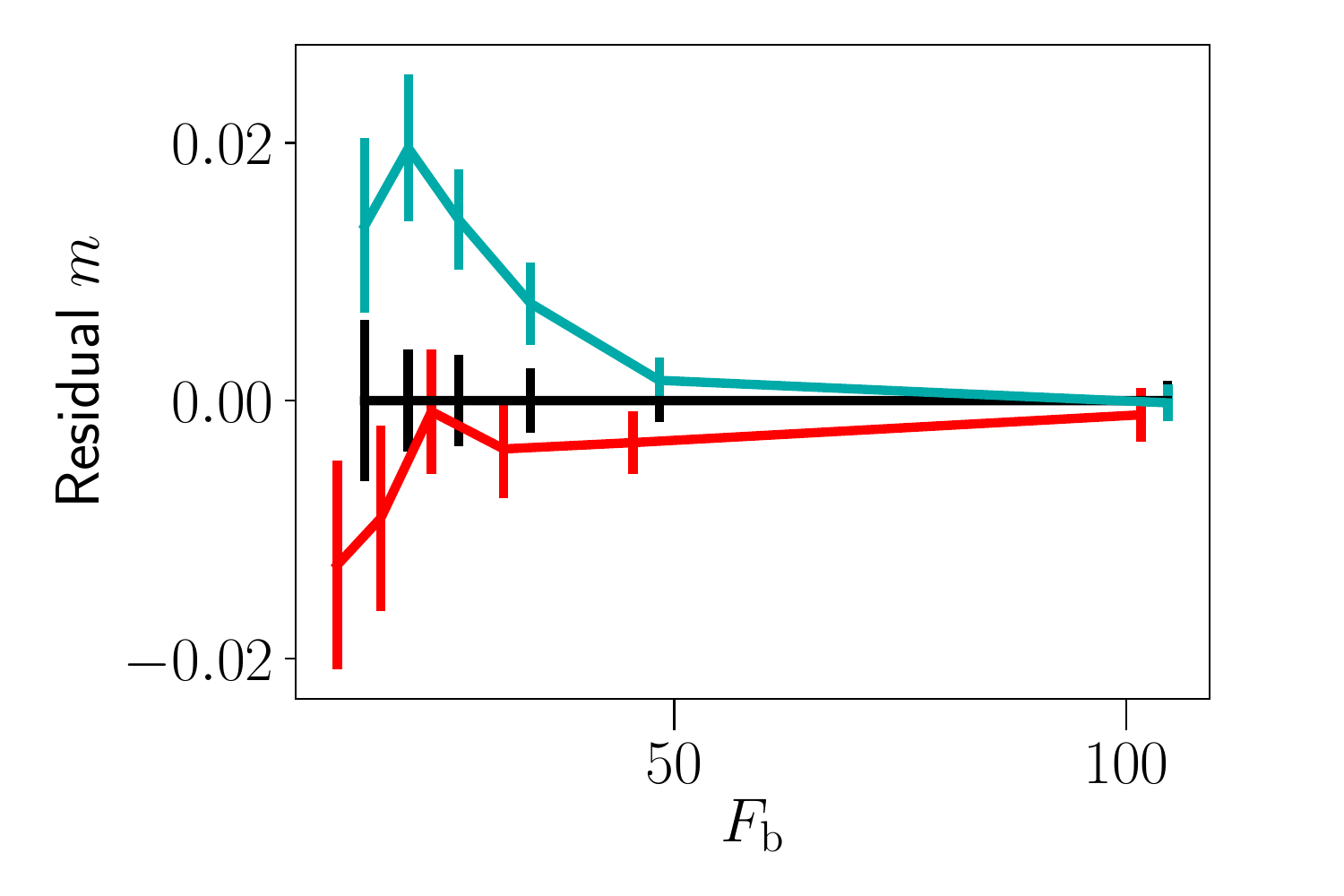}
\includegraphics[width=.48\linewidth]{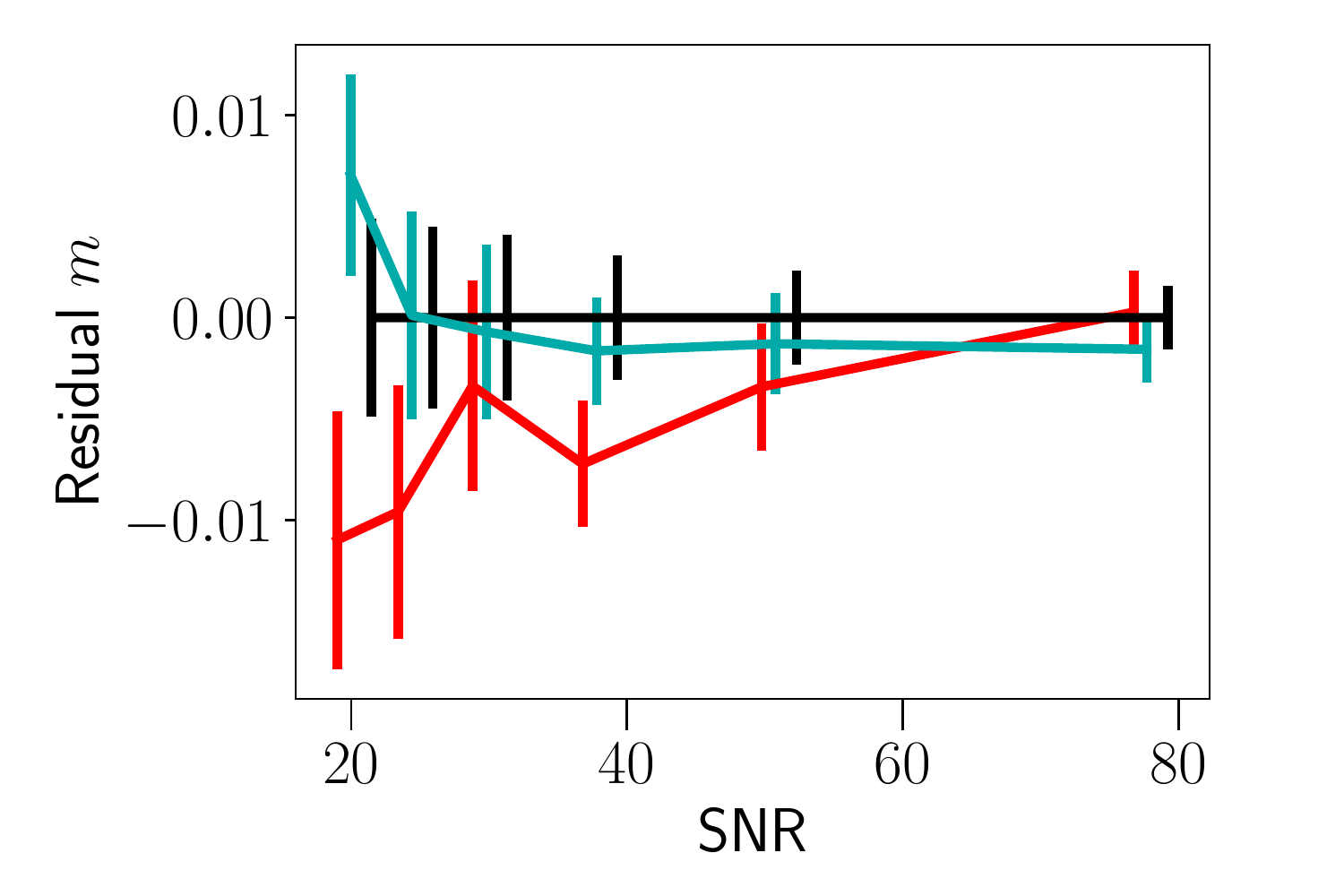}
\caption{Residual multiplicative shear bias as a function of input galaxy properties: disc flux (top left), disc half-light radius (top right), bulge flux (bottom left), and S/N (bottom right). Red lines show the calibration from \metacal. The black lines show the calibration using the true bias, and the cyan lines show the calibration using the bias estimate from NNSC.}
\label{fig:cal_m_reg}
\end{figure*}

\subsection{2D dependencies}\label{sec:2d_dependencies}

Fig. \ref{fig:2d_m_comp}  shows examples of the multiplicative shear bias 2D dependences. Here the multiplicative bias $m_1$ is represented in colour, the left panels show the true bias, and the right panels the estimates from NNSC. In this case, the top four panels show the dependences as a function of input simulation parameters (not accessible for NNSC), and the bottom panels show the dependences on measured parameters used for the training. NNSC clearly predicts the shear bias as a function of combinations of two input properties well. As before, the method was trained to describe shear bias as a function of the measured properties, in consistency with the good performance in the bottom panels, but the predictions on the input simulation parameters depend on how strongly these properties are constrained by the measured parameters. The method describes shear bias as a function of shape parameters very well (middle panels), but it underestimates the values for some galaxies with a very low Sérsic index $n$ and intermediate radius because $n$ was not estimated and no properties referring to this parameter were used for the training. The performance of the model would improve when more measured properties on the training that are correlated with $n$ were used (e.g. a fitting parameter estimation of the galaxy profile).

\subsection{Residual bias}

In order to test the performance of the shear calibrations, we analysed the residual bias estimated from a linear fit of Eq. \ref{eq:g_relation} after the galaxy samples were corrected for their bias. Here we include \metacal\ as a reference for an advanced shear calibration method so that we can compare our performance with currently used approaches. With this we do not aim to show a competitive comparison of the methods, but to confirm the consistency of NNSC with respect to what can be expected for a reliable method. The two methods are intrinsically different and affected by different systematics, therefore a combination of the two methods can be a very complementary and robust approach for scientific analyses. Moreover, the two methods can be differently optimised for the performance in different types of data, and here we do not pretend to show the best-case scenario for any of them. For details of the implementation done for \metacal,\ see Appendix \ref{sec:metacal}.

In Fig. \ref{fig:cal_m_reg} we show the residual multiplicative bias $m$ as a function of several input properties found for three different approaches. In black, the calibration was made using the true shear bias obtained from the image simulations. This represents the best-case scenario where the shear bias has been perfectly estimated and gives an estimate of the statistical uncertainty of the measurement.
The red and cyan lines show the residual biases from \metacal\ and NNSC, respectively.

Both methods show a residual bias $\text{of less than 1\%}$  for most of the cases, and the performance depends on the galaxy populations. In general, very good performances are found for both methods for bright or large galaxies. In the case of \metacal, the residual multiplicative bias increases to $\text{about } 2\%$  for small and dim galaxies, showing that the sensitivity of the method depends on the signal of the image, as expected. For NNSC, the performance depends on the explored property; it extends from negligible residual bias for any S/N to a residual bias of up to $4\%$  for galaxies with a very dim disc. We recall that these input properties from the simulations that were not used for the training, therefore the performance of the calibration depends on the correlation between the measured properties and these input properties. Because we use an estimate of the S/N in the training, the performance of the calibration is excellent even for galaxies with a very low S/N. On the other hand, galaxies with a low disc flux are not well characterised from the measured properties we used in the training. They can be a combination of dim galaxies and galaxies with a very small disc fraction, and no measured properties aim to describe the morphology of the disc regardless of the contribution from the bulge. Because of this, the NNSC performance on those galaxies is worse. However, in real data applications we will never have access to this input information, and the galaxies will always be selected from measured properties that can be included in the training set, so that this problem will not appear on real data applications as it does here. Instead, this will produce selection effects that are discussed in Sect.~\ref{sec:selection_bias}.

\subsection{Robustness with realistic images}\label{sec:realistic_imgs}

\begin{figure}
\centering
\includegraphics[width=.98\linewidth]{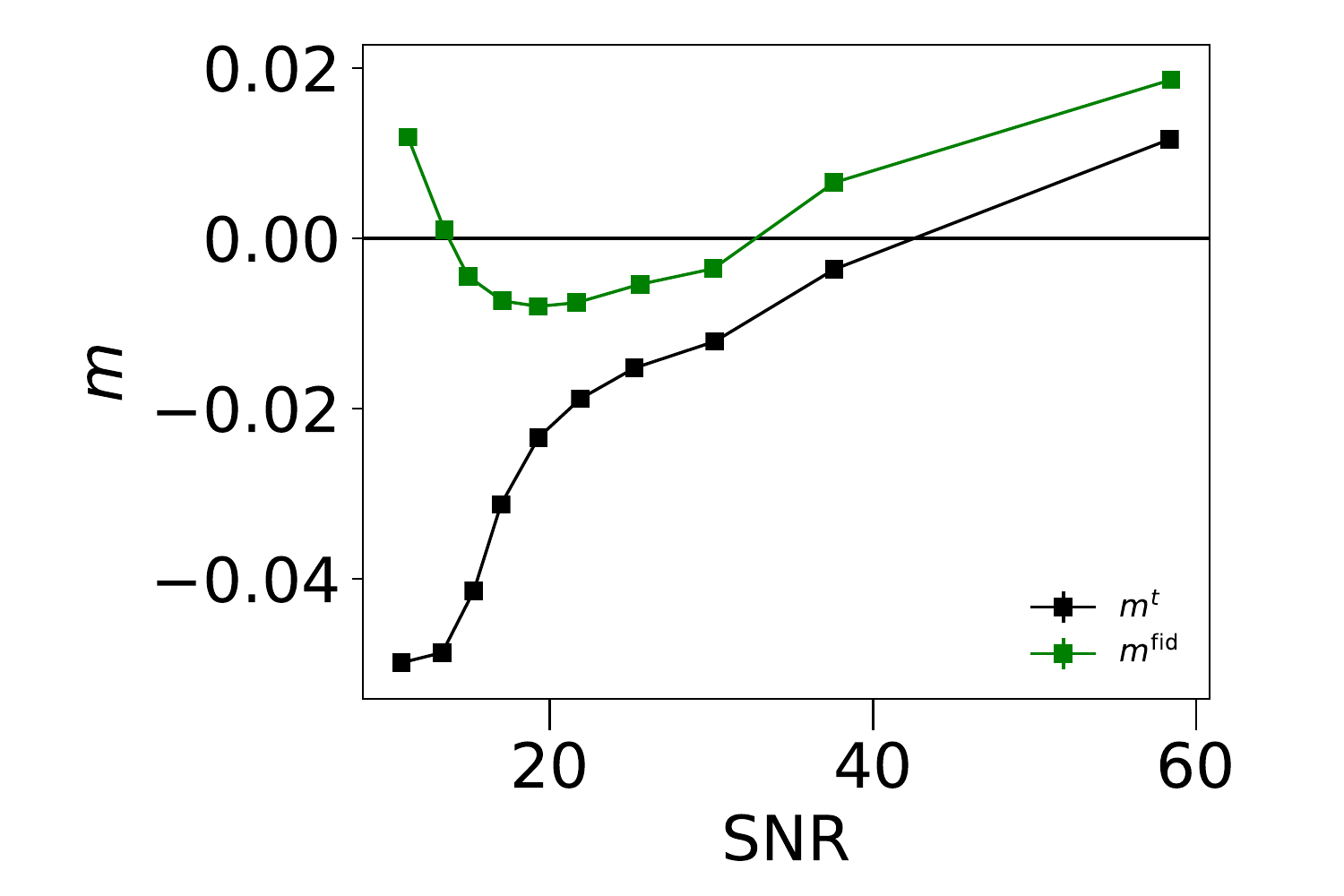}
\includegraphics[width=.98\linewidth]{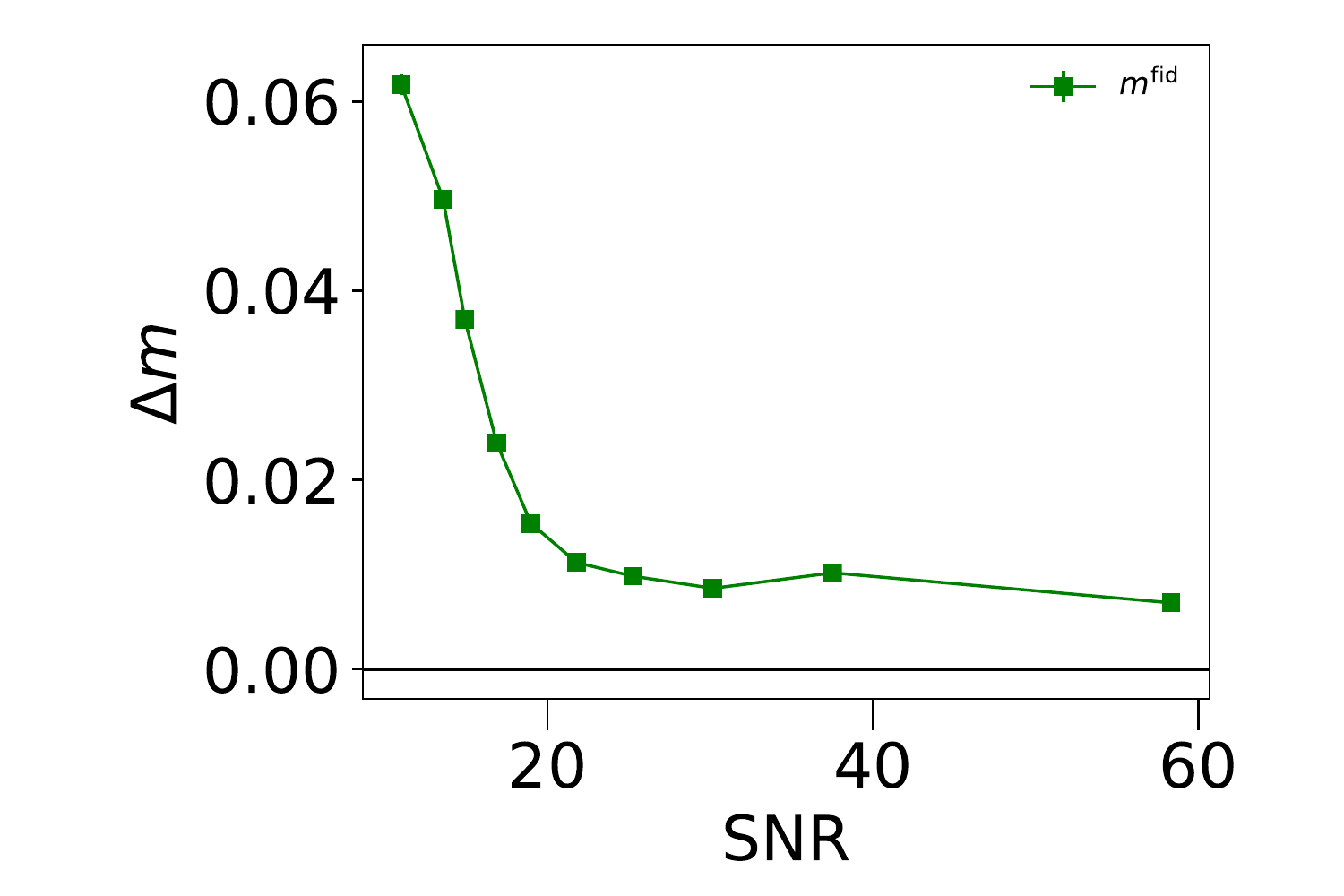}
\caption{Multiplicative bias predictions for GREAT3-RSC images as a function of S/N. The top panel shows the bias estimate from NNSC (green line) compared to the true bias (black line). The bottom panel shows the difference between the estimated and true bias as a function of S/N. }
\label{fig:rsc_perf}
\end{figure}

\begin{figure}
\centering
\includegraphics[width=.98\linewidth]{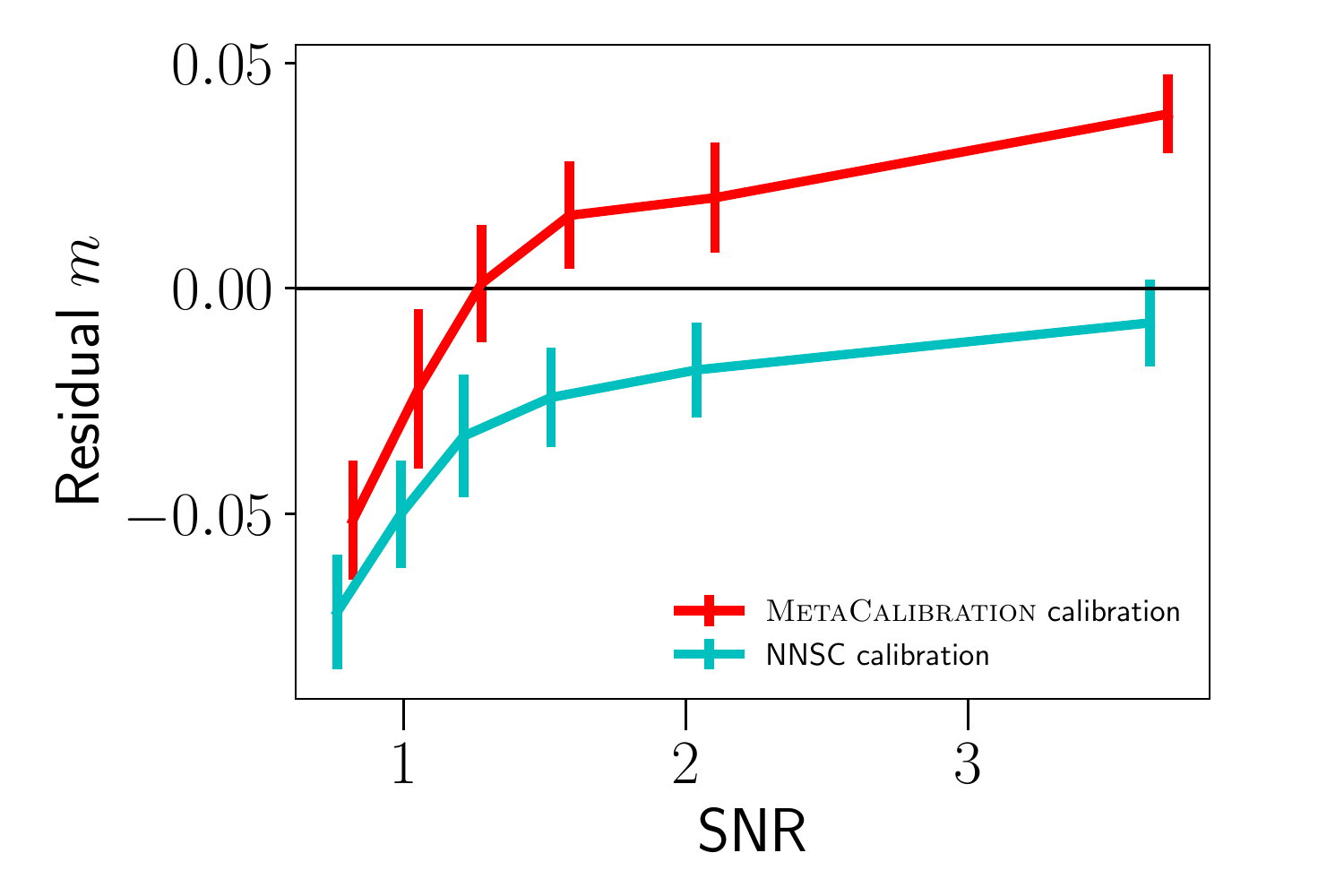}
\caption{Residual multiplicative bias after calibration for realistic images using \metacal\ (red line) and
the calibration from the bias estimate of NNSC (cyan line).}
\label{fig:res_m_rsc}
\end{figure}

The NNSC model has been trained with a specific set of image simulations based on the CSC branch from GREAT3. To evaluate the potential effect of applying this model to real data with no further training, we used the NNSC model that was trained with the GREAT3-CSC images but applied to calibrate GREAT3-RSC images instead. In Fig.~\ref{fig:rsc_perf} we show the estimated bias compared to the real bias obtained from sheared versions of the images as in \cite{Pujol2018} using equations \ref{eq:mean_response}
 and \ref{eq:ind_add_bias}. The top panel shows the dependence of $m$ on S/N, the bottom panels shows the error on the estimation of $m$. An error of up to $6\%$  for the lowest S/N and of $\text{about } 1\%$  for galaxies with S/N$>20$ is evident. This indicates that the model trained on analytic, simpler galaxy images does not yield perfect estimate for realistic galaxy images, but it still obtains errors of $\text{about } 1\%$  for the majority of the cases.

 Fig.~\ref{fig:res_m_rsc} shows the residual bias after the calibration was applied to the GREAT3-RSC images for NNSC and \metacal. Again, the training for the NNSC calibration was that of the CSC branch. The difference in S/N values with respect to previous figures is that now we use the S/N estimate from \ksb\ (we used the input parameter for \galsim\ or the S/N from \cgfit\ for the remainder). For NNSC a residual bias is visible that decreases with S/N, consistent with the bias estimates from Fig.~\ref{fig:rsc_perf}. It is beyond  the scope of the paper to improve this calibration by applying a refinement to these images because we wish to show, on the one hand, the potential performance of the method (Sects.~\ref{sec:bias_preds}-\ref{sec:2d_dependencies}), and on the other hand, the effect of applying a crude calibration to a realistic data set for which the model was not trained (this section). Otherwise, an easy improvement could be achieved with a refinement of the model on realistic galaxy images, or by identifying poorly estimated objects (we found that the main contribution of this error comes from objects whose shear responses were estimated outside the range of values represented by the training, indicating a misrepresentation of these objects).

 \metacal\ also shows a significant residual bias dependence on S/N for GREAT3-RSC images. Although the residual bias over the entire population is very weak and consistent with previous analyses \citep{Huff2017}, the method shows a negative residual bias for galaxies with a low S/N and a positive one for galaxies with a  high S/N. We found this to be specific for the RSC images and this particular implementation. Different from CSC images, the estimated shear responses of \metacal\ show a weak dependence on S/N. This is caused by some images that our \ksb\ implementation interprets to be small, and a small window function was applied to them for the shape measurement. This produces a very similar calibration factor ($\sim 5\%$ positive) for all S/N values, producing a $\text{shift of about } 5\%$ in the residual bias. We ignored the origin of this low sensitivity; it can come from a combination of factors. First, RSC images are created from pixelated and noisy real images that have been deconvolved with their PSF to which then a shear was applied, making these images imperfect. In additiont, \metacal\ is ran, which again modified the images with a deconvolution, shearing, re-convolution, and a noise addition. Finally, \ksb\ estimates the optimal size of the window function to estimate the galaxy shape.

 \section{Discussion}\label{sec:discussion}

 \subsection{Potential limitations and solutions for NNSC}

 \subsubsection{Selection bias}\label{sec:selection_bias}

 The aim of this paper is to show the performance of NNSC in calibrating shear measurement bias. To this end, we applied the following catalogue selection in order to remove any selection bias from the data. Any galaxy whose detection or shape measurement failed in any of the processes was removed from the catalogue, together with all its shear versions. This included not only the shear versions that were simulated with \galsim,\ but also the images derived from the \metacal\ processing so that \metacal\ has no selection bias either. Moreover, when the data were split into bins of a measured variable, all the shear
versions were removed as well  if a galaxy fell in different bins for different shear versions. With this, the selected data of our results were identical for all shear versions, and selection bias was forced to be zero.

 This procedure can be applied here for the purpose of presenting a method in simulations, but in real data selection, effects cannot be avoided and need to be calibrated. In particular, selection bias comes from the fact that the selection function depends on the shear and is usually of the same order of magnitude as shear measurement bias \citep{Conti2016,Mandelbaum2018b}. \metacal\ takes into account the shear dependence of selection effects and also calibrates selection bias in a similar procedure as it does for measurement bias, as described in \cite{Sheldon2017}. \cite{Sheldon2019} also explored a calibration of selection and measurement bias simultaneously in the presence of blended objects.

 Analogously, NNSC can potentially be used to also calibrate selection bias. This would involve applying the same selection process to all galaxies independently of the shear and measuring the shear responses to this selection, as described in \cite{Sheldon2017} and in Sect.~7.2 of PKSB19:
 \begin{equation}
   \left\langle R_{\alpha\beta} \right\rangle \approx
   \frac{\left\langle e^{\rm obs, +}_\alpha \right\rangle - \left\langle e^{\rm obs, -}_\alpha \right\rangle}{2 \Delta g_\beta} ,
   \label{eq:ind_bias_estim_mean}
 \end{equation}
 where the ellipticities are measured for the case with no shear, and the $+$ and $-$ superscripts refer to the applied selection, corresponding to the catalogues obtained from the positive and negative shear versions, respectively. A supervised training could be applied to learn the shear response on selection. This would involve adapting the method so that the selection is specified in the input data (e.g. with weights specifying the selection for each shear version). Then the cost function involves the average selection shear response over a subset of the catalogue, as

 \begin{equation}
 C = \frac{1}{b_\textrm{s}} \sum_{i=0}^{b_\textrm{s}} \sum_{\alpha = 1}^2 \sum_{\beta = 1}^2 (\frac{ w^+_i e^{\rm obs}_{i,\alpha} -  w^-_i e^{\rm obs}_{i,\alpha} }{2 \Delta g_\beta} - R_{i,\alpha \beta}^\textrm{e})^2,
 \end{equation}

 where $w^{+,-}_i$ specifies the selection of galaxy $i$ for the cases with positive or negative shear (it can be a weight from $0$ for undetected cases to $1$ for detected cases with the full signal), $e^{\rm obs}_{i,\alpha}$ is the observed ellipticity for the case with no shear, and $R_{i,\alpha \beta}^\textrm{e}$ is the output estimated shear response of the training.

 \subsubsection{Model bias}

 \begin{figure*}
 \centering
 \includegraphics[width=.48\linewidth]{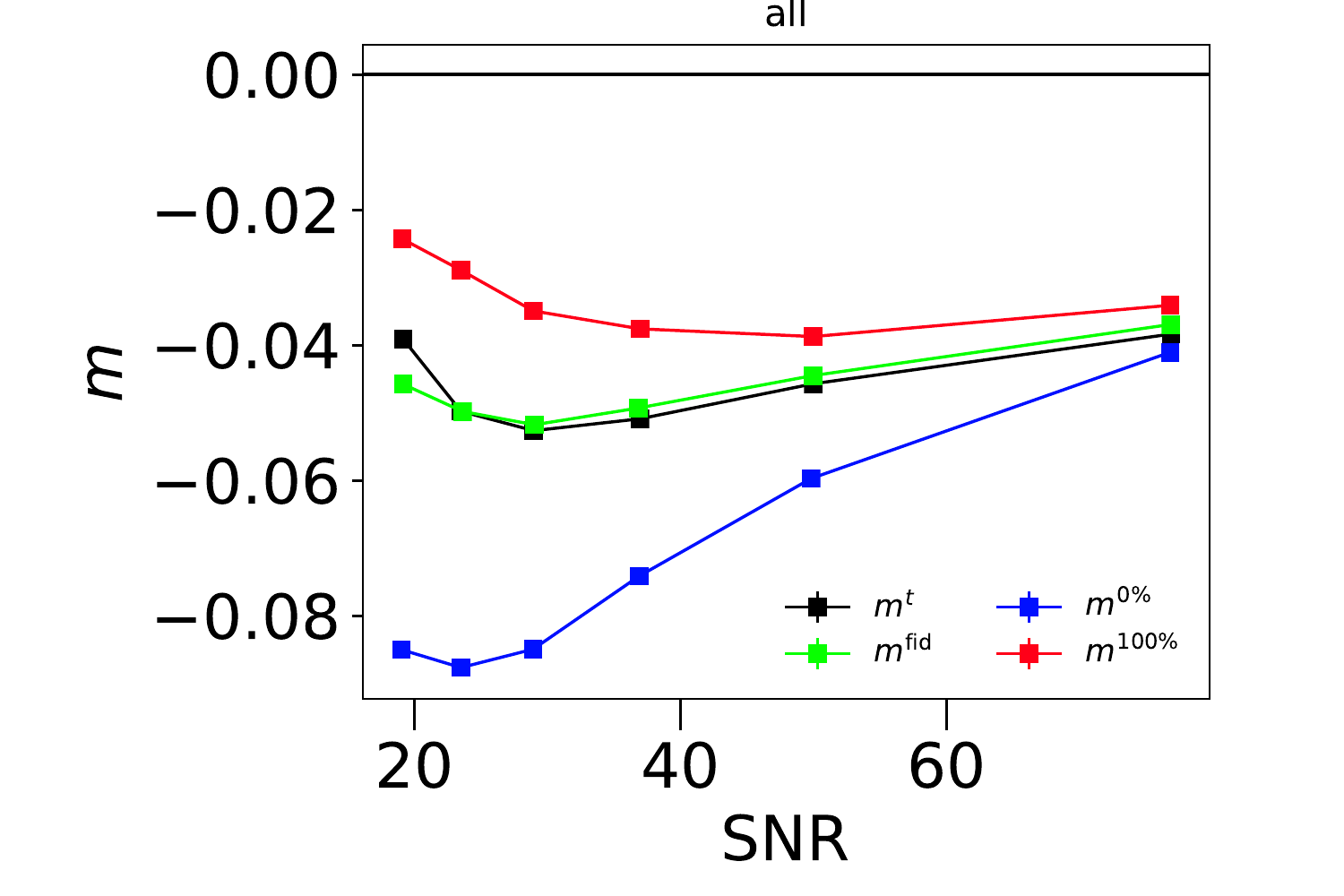}
 \includegraphics[width=.48\linewidth]{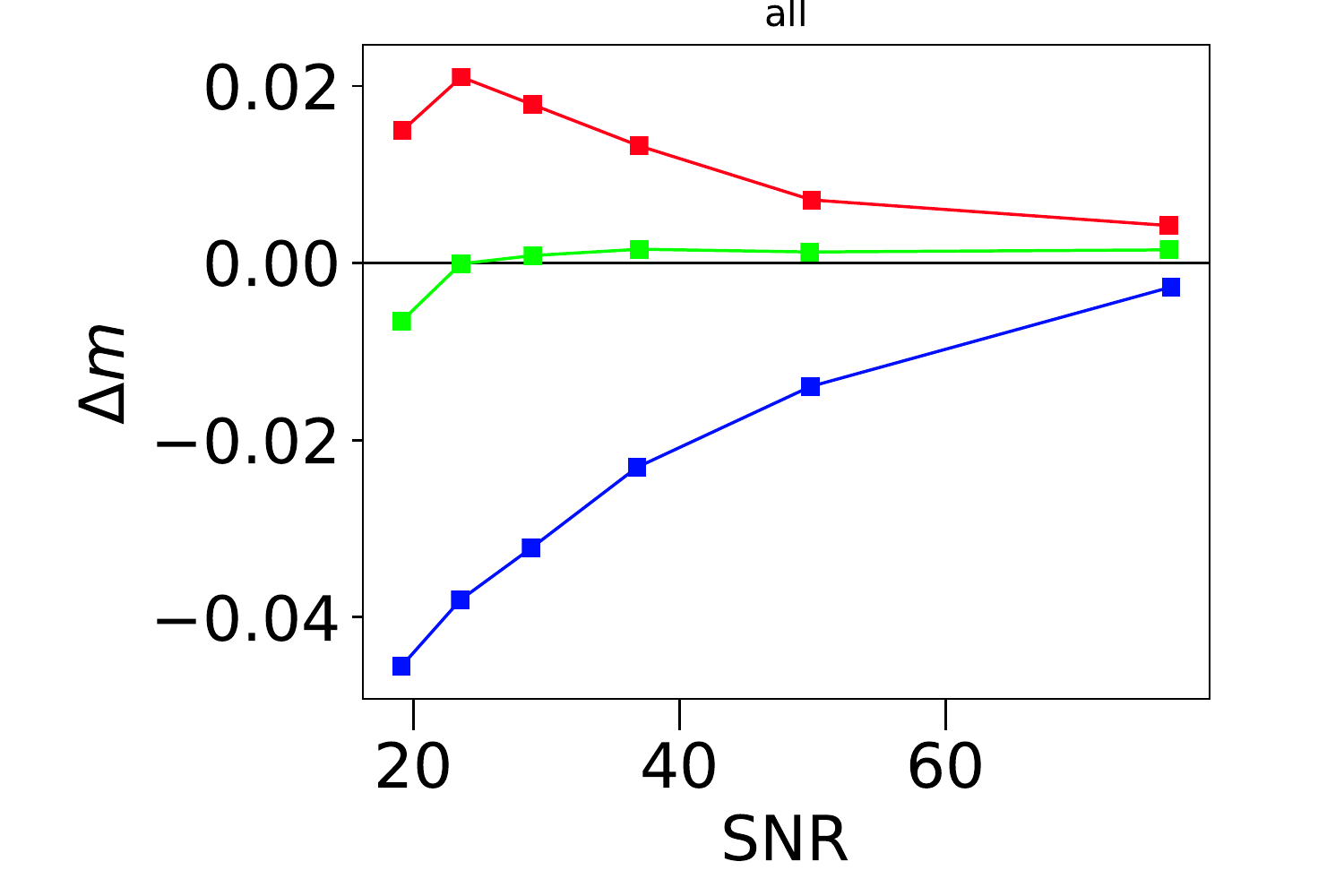}
 \includegraphics[width=.48\linewidth]{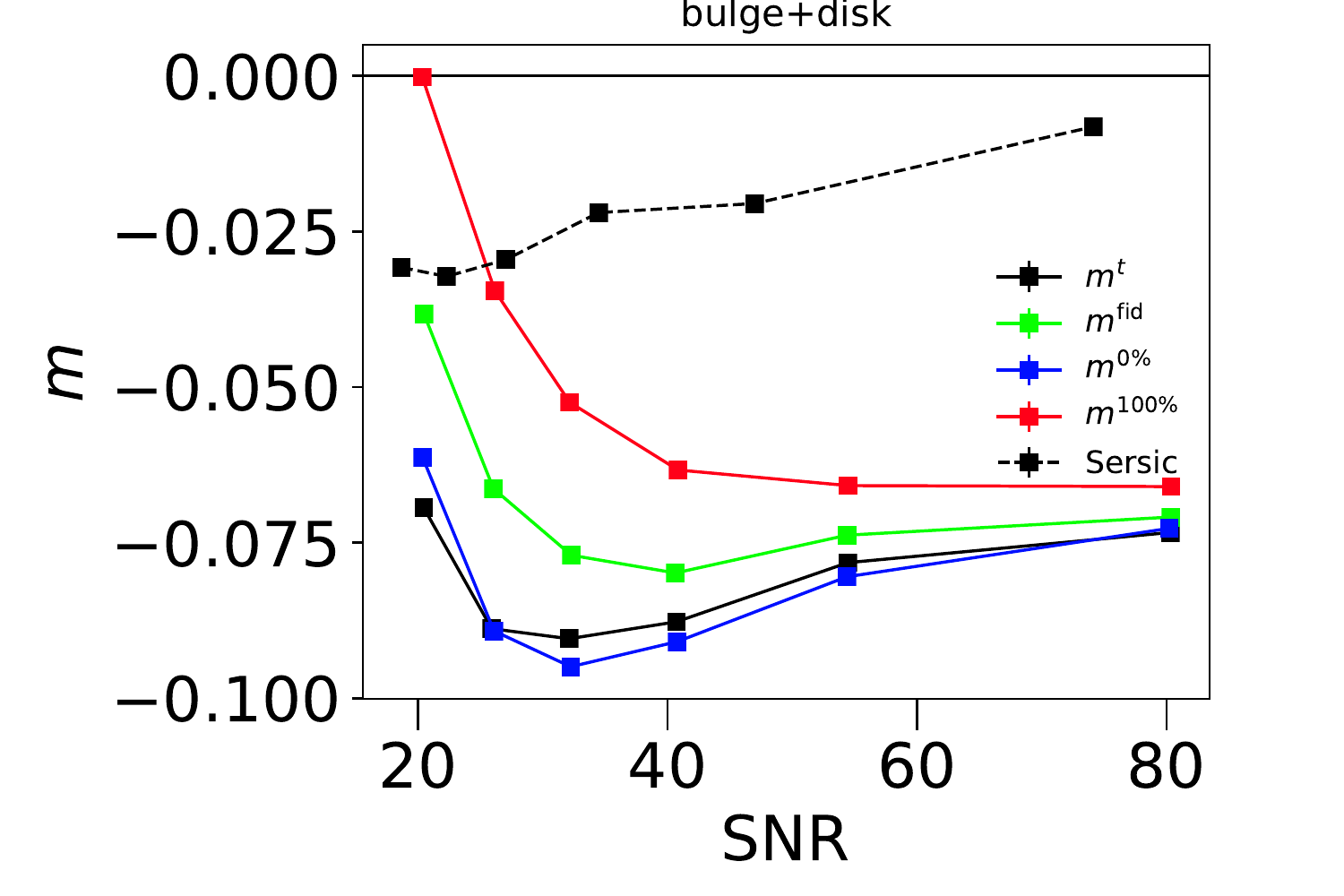}
 \includegraphics[width=.48\linewidth]{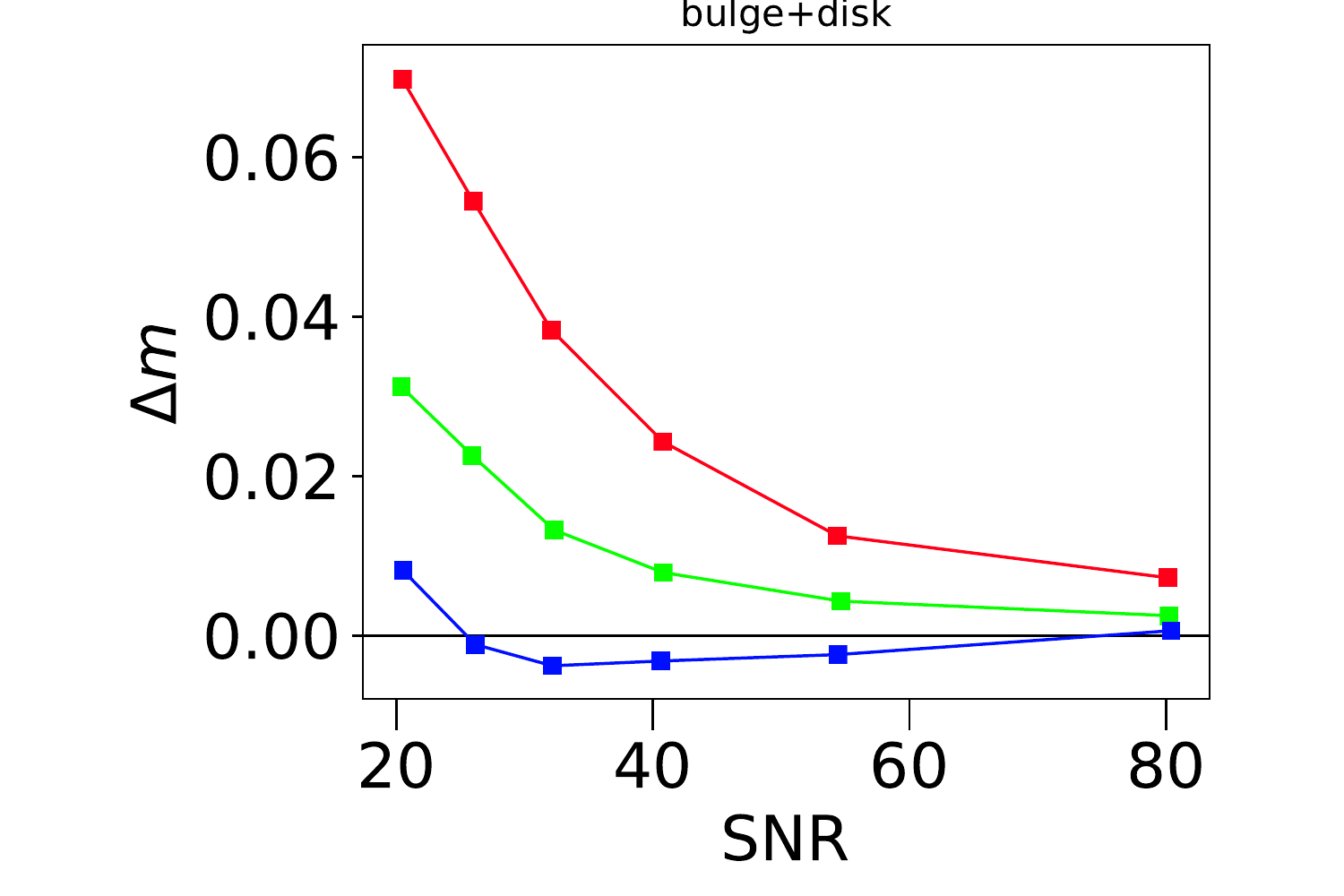}
 \includegraphics[width=.48\linewidth]{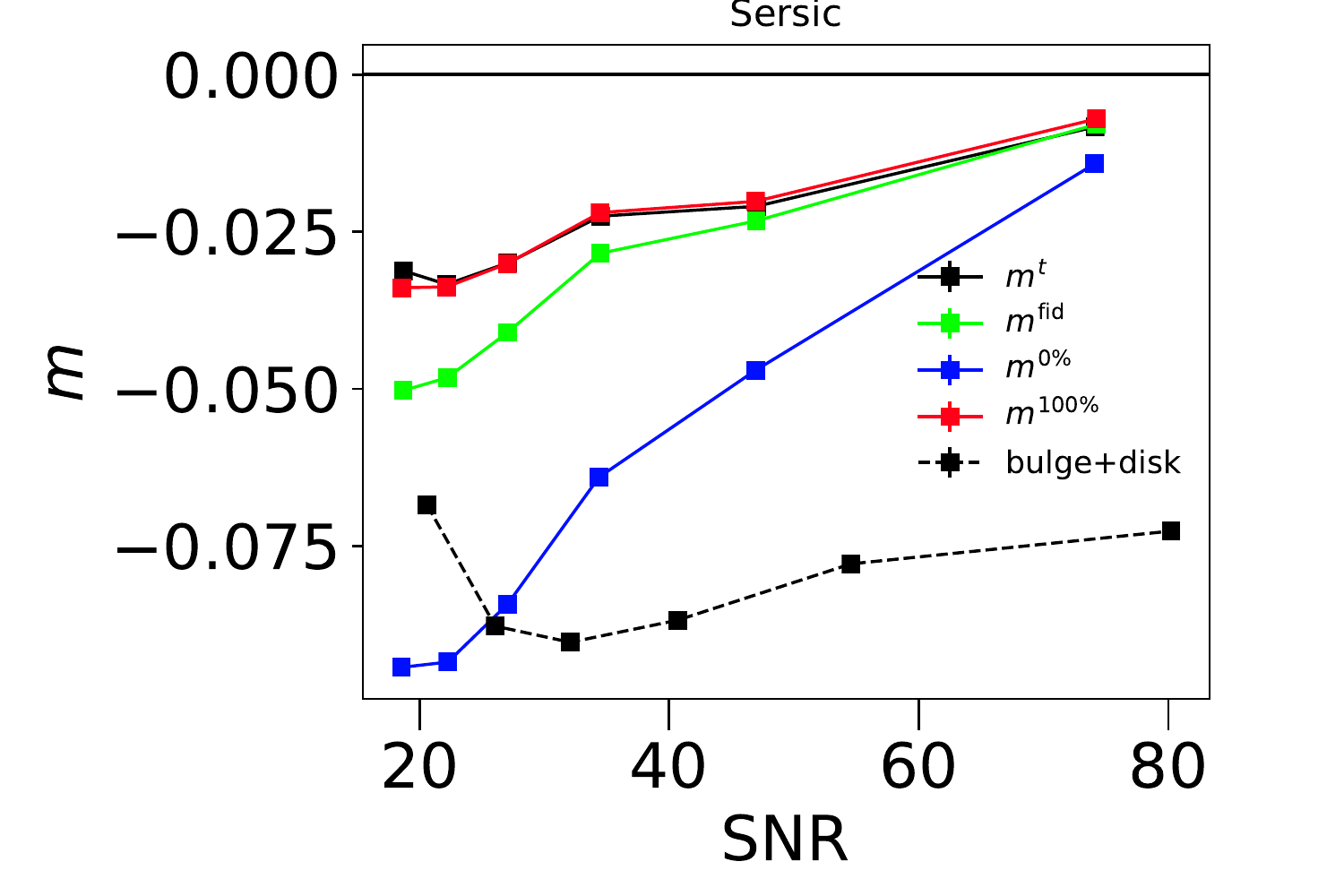}
 \includegraphics[width=.48\linewidth]{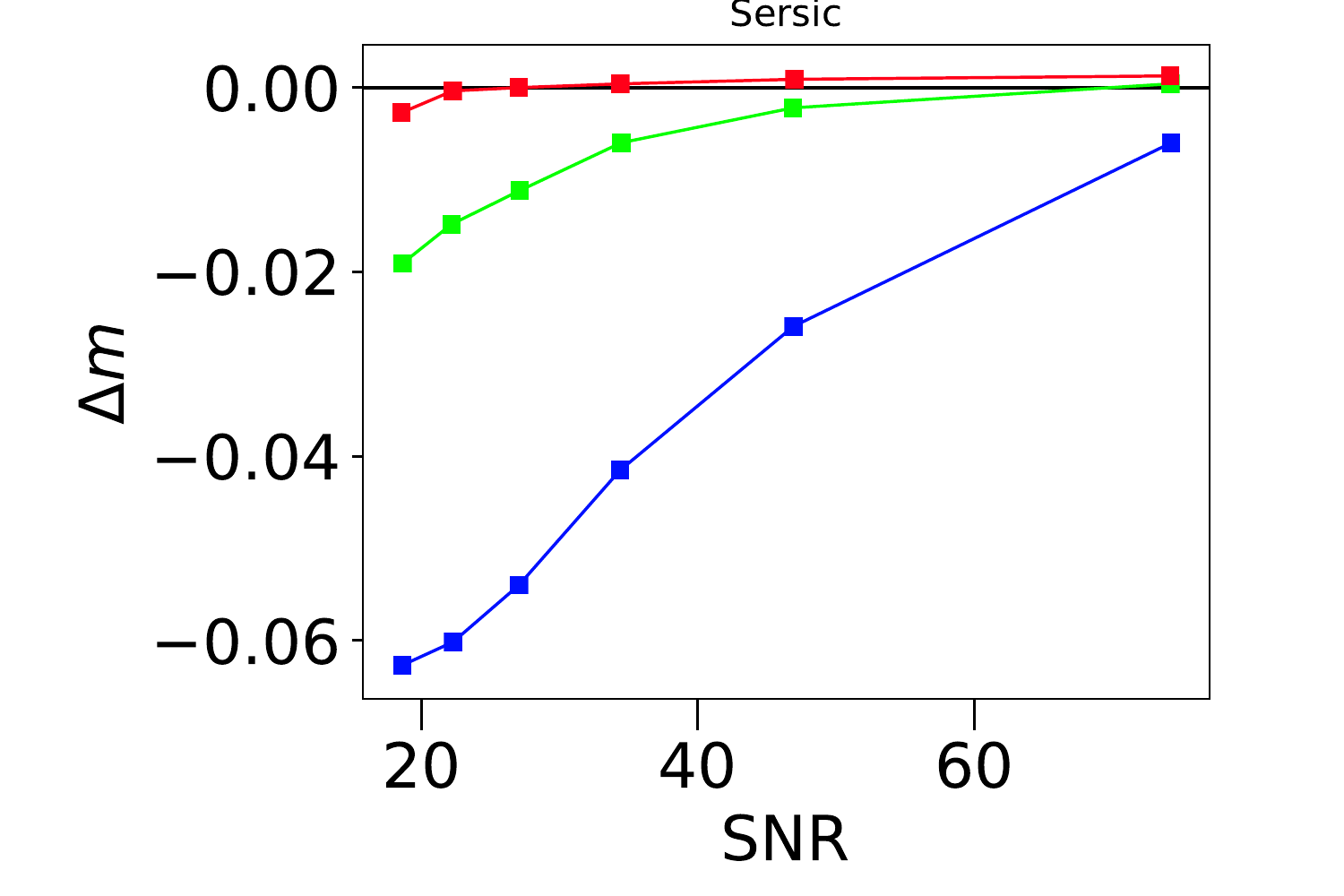}
 \caption{Shear bias predictions using only single Sérsic (in red), only bulge+disc (in blue), or the real population (green) tested on the whole population (top), on only disc+bulge galaxies (middle) and on single Sérsic galaxies (bottom). Solid lines show the true bias of the populations, and the dashed black lines show the shear bias of the excluded population (the true bias for single Sérsic galaxies in the middle panels and for bulge+disc in the bottom panel).}
 \label{fig:model_bias}
 \end{figure*}

 \begin{figure*}
 \centering
 \includegraphics[width=.48\linewidth]{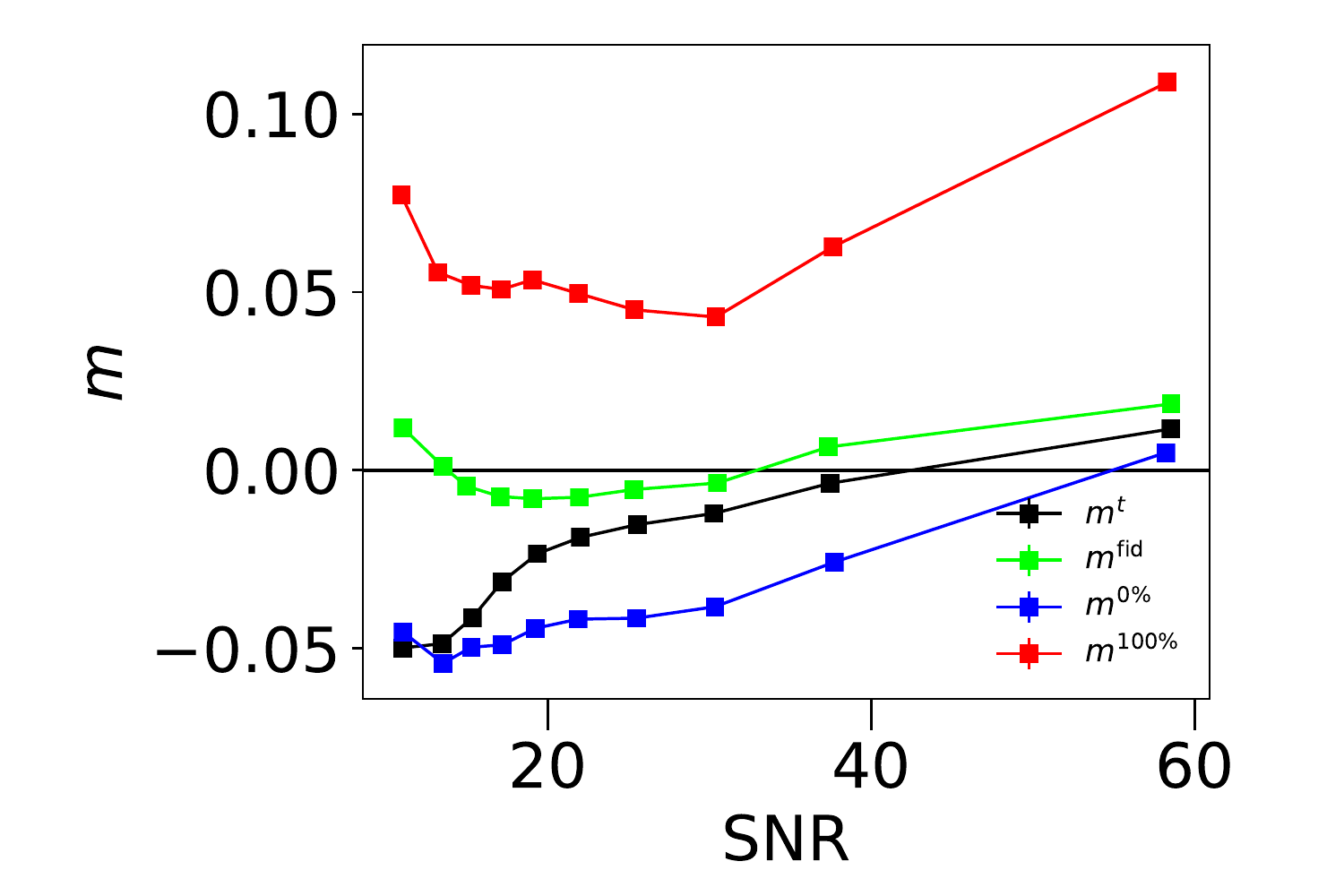}
 \includegraphics[width=.48\linewidth]{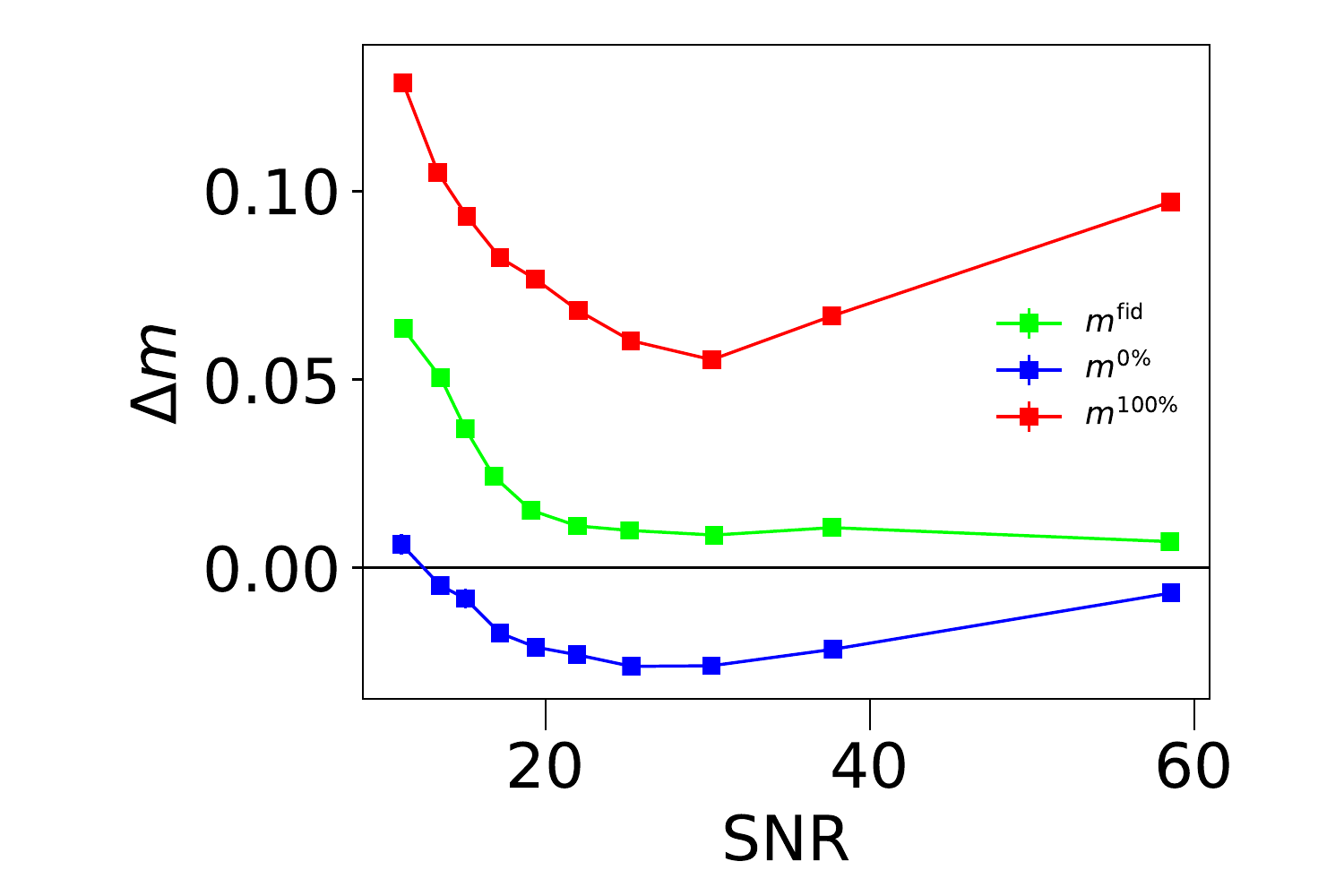}
 \caption{Same as in Fig.~\ref{fig:model_bias}, but applied to the realistic RSC images. }
 \label{fig:rsc_model_bias}
 \end{figure*}

 For the results of our method in this paper, we used the same type of population for the training process as for the test and calibration. However, it is known that shear bias depends on the galaxy profile models (known as model bias). The images used in this paper consist of a mix of single Sérsic galaxies and galaxies with the sum of a bulge (de Vacouleurs) and a disc (exponential). For the original training, testing, and calibration, we used a population in which $61\%$ of the galaxies have single Sérsic profiles (this corresponds to the fraction in the whole set of images).

 Here we quantify the effect of the model bias on our method that comes from these two different models by testing the performance of the method when different single Sérsic fractions are used for the training and the testing steps. In Fig. \ref{fig:model_bias} we show the performance of the estimated multiplicative bias using different Sérsic population fractions for the training data as specified in the legends ($m^{\rm fid}$ corresponds to the original fraction of $61\%$, and $m^t$ shows the true bias). The top panels show the results applied to the original population with a Sérsic fraction of $61\%$, in the middle panels we show the results on galaxies with a bulge and a disc (Sérsic fraction of $0\%$), and in the bottom panels we show the results applied to Sérsic galaxies (Sérsic fraction of $100\%$). The left panels show the bias dependence, and the differences with respect to the true are shown in the right panels.

 In all the cases, the best performance is obtained when the training population coincides with the test population. On the other hand, all the cases give different shear bias predictions for the different test populations. For example, the model trained with only Sérsic galaxies predicts $m \sim -0.06$ for the galaxies with a disc and a bulge and $m \sim -0.02$ for the single Sérsic galaxies. Although the true $m$ changes a $5-6\%$ between the two populations, training with the single Sérsic population alone gives $\sim 1\%$ error on the other population. In the opposite case, the model trained with disc and bulge galaxies predicts $m \sim 0.08$ for the same population and $m \sim 0.02-0.08$ for the single Sérsic population. This means that all the predictions are sensitive to the differences between the different populations, although the model bias still remains non-zero. This means that we have captured only a part of the dependency of the bias on S/N and type of galaxy. The red lines in the middle panel and the light green lines in the bottom panels show the extreme cases when the models were trained with a completely different population, and they give a model bias of $\sim 1\%$ for bright galaxies.
  Possible ways to reduce this model bias could be a) training with more complex models, b) using more input measured  complementary properties that can help to increase the sensibility to more complex images, and c) using convolutional neural networks (CNNs) to exploit the information at the pixel level.

 Finally, in Fig.~\ref{fig:rsc_model_bias} we show the same bias predictions, but now applied to RSC images. This shows the effect of model bias when different analytical models are trained and applied to realistic images. The fiducial model gives better predictions on RSC images than the extreme cases where the training is only made with one galaxy model, at least for S/N$>20$. A good performance is encouraging for the practical effect of model bias in real observations with our training, and using more sophisticated or realistic models for image simulations would potentially improve the performance, as discussed in Sect.~\ref{sec:realistic_imgs}.

 \subsection{Advantages of the NNSC method}\label{sec:advantages}

 We presented NNSC as a new method for shear calibration. The method is different from others such as \metacal, self-calibration methods, the commonly applied calibration from shear bias measurements in binned properties from simulations or other ML approaches. We do not claim one method or approach to be better than the other, but we highlight several strengths of our method.

 First, NNSC has the advantage that it can obtain a good performance in a matter of hours with a few thousand images, which is a very efficient ML approach for shear calibration. This is because, on the one hand, the input data are a reduced set of measured properties that significantly reduce the architecture dimensionality (compared to other ML approaches such as convolutional neural networks), and on the other hand, by focusing on minimising the error on the estimated individual shear response, we avoided the shape noise contribution of the intrinsic ellipticity (see PKSB19 for a detailed discussion of this contribution). It could also be possible to build an estimate of the shear bias straight from the galaxy and PSF images (e.g. from aCNN), but this would require a much more complex network architecture (e.g. accounting for the effect of the PSF, galaxy morphology, etc.). Learning from image properties already reduces the complexity of the images, without losing too much information of the shear bias.

 As an example of an ML approach, \cite{Tewes2018} minimised the residual bias over the average measured ellipticities so that the output wass an unbiased shear estimator. Because of the intrinsic ellipticity contribution, the method requires a very large catalogue ($10^6 - 10^7$ objects) and uses a batch size of $500,000$ objects (distributed so that the intrinsic ellipticity cancels out) to ensure that in each minimisation step, the shape noise from the intrinsic ellipticity is low. One suggestion to improve the computational performance of \cite{Tewes2018} would be to minimise the individual responses  (using  sheared versions of the same images as in this paper and in PKSB19) instead of the residual bias over average ellipticities.

 In another approach, \cite{2010ApJ...720..639G} minimised shear bias for a \ksb\ estimator by training the ellipticity measurement errors using the original measurements from \ksb,\, the flux measured from \pse,\ and some tensors involved in the shape measurement process. The approach is similar to ours in the sense that they considered a set of properties to estimate errors on the shape measurements, but our method directly estimates the shear response, which avoids shape noise. As in \cite{Tewes2018}, \cite{2010ApJ...720..639G} used $\sim 10^7$ objects for the training sets.

 Another potential of the NNSC method is that it can be easily implemented for any survey for which we have image simulations (this is required for a good validation of the survey exploitation). To apply NNSC, we only need to produce copies of the same image simulations with different shear values applied, so that we can obtain individual shear responses. The set of measured properties to be used for the training is arbitrary and can be chosen according to the interests and the pipeline output of the surveys. Even a simple application using a few properties of interest will already be an improvement with respect to common calibrations obtained from shear bias measurements in simulations as a function of two properties only. Moreover, as described in Sect.~\ref{sec:selection_bias}, the method can be extended to also calibrate selection bias.

 Finally, the method allows us to use a large set of measured properties as input. When properly trained, this allows the calibration to be more reliable than the particular population distribution of the training with respect to the real data. If the bias dependences on many properties are correctly learnt, the particular distribution of the population over these properties should not affect the performance of the calibration (provided that no other unaccounted-for properties affect shear bias and that the simulated population is representative of the real data).

 \subsection{Importance of using complementary methods}

 We have used two different methods (\metacal\ and NNSC) and analysed their performances in image simulations. Both methods perform well, and they are complementary because they do not share the same systematics. Different implementations of the \metacal\ pipeline, as well as using other shape estimators, might improve or better adapt to the particular data. As for our model, the scope of this paper is not to show the best implementation of \metacal\ for these particular data, but to include it to compare NNSC with an advanced modern code.

 Because the two models are complementary, using both for the same scientific analyses is a more suited approach to ensure the  reliability of the results. For this reason, we encourage researchers to include at least two independent shear measurements and calibration methods for scientific analyses on galaxy surveys, as was done in Dark Energy Survey \citep{Jarvis2015,Zuntz2017}.
 This allows us to better identify systematics from the discrepancies between the methods that otherwise might be missed. At the same time, consistent results from two different and independent methods always give reliability to the scientific results, a crucial aspect for future precision cosmology. The combination of NNSC and \metacal\ brings a good complementarity because it uses an ML approach based on measurements from image simulations and a method that is independent of image simulations, but limited by other numerical processes.

 \section{Conclusions}\label{sec:conclusions}

 Machine learning is a promising and emerging tool for astronomical analyses because it can characterise complex systems from large data sets. It is then especially well suited for shear calibration, where many systematics complicate the behaviour of shear bias and its calibration.

 We presented a new shear calibration method based on ML that we call neural network shear correction (NNSC). We have also made the code publicly available. The method is based on galaxy image simulations that are produced several times with different shear versions, but the remaining conditions are preserved. With this, we obtained the individual shear response of the objects that served us for a supervised ML algorithm for estimating the shear responses of objects from an arbitrarily large set of measured properties (S/N, size, flux, ellipticity, PSF properties, etc.) through a regression approach.

 This ML approach allowed us to characterise the complexity of shear bias dependences as a function of many properties, a complexity that we explored in \cite{Pujol2017}. The advantage of ML is that it is an especially well suited approach to reproducing very complex systems so that we can include a large set of properties on which shear bias can depend. With these, the algorithm identifies the contribution of each of the properties and their correlations to estimate shear bias. With the individual shear bias estimates of galaxies, we can then apply a shear calibration based on the average statistics of shape measurements as in many other shear calibration approaches.

 We used image simulations based on the GREAT3 \citep{Mandelbaum2014} control-space-constant branch to explore the method, apply the training algorithm of NNSC, and evaluate its performance through tests and validations. We obtained a performance beyond Euclid requirements ($\Delta m_1 = (4.9 \pm  1.1) \times 10^{-4}$, $\Delta m_2 = (0.0 \pm  1.1) \times 10^{-4}$, $\Delta c_1 = (-3.1 \pm 0.7) \times 10^{-4}$ and $\Delta c_2 = (1.6 \pm 0.7) \times 10^{-4}$) for the estimates of the average shear biases, and we showed shear bias dependences on one and two properties below the 1\%\ error.  This performance was achieved with only $\sim 15$ CPU training hours using $128,000$ objects, which is a very cheap and fast training compared to common ML approaches (the method from \cite{Tewes2018} takes about two  CPU days with $10^6 - 10^7$ objects). This means that the NNSC approach has great potential, but is also very easy to apply to galaxy surveys because it only requires different shear versions of the same image simulations and low computational and storage capacities.

 We compared the residual bias after calibration as a function of several input properties with the advanced method \metacal. This resulted in similar performances and showed different dependences because of the differences between the approaches and their systematics.

 We quantified the effect of model bias by applying the calibration to different galaxy morphologies than were used for the training. We found  errors of a few percent in some extreme cases that could be improved by refining the training with a more proper or sophisticated data set. Although it is not developed in this paper, the method can be easily adapted to also learn selection bias and calibrate for it just by adding information about the data selection as input data for the training and applying small changes in the cost function.

 Our method is an implementation of ML based on a simple DNN architecture leading to fast calibration that can be easily applied to weak-lensing analyses of current and future galaxy surveys and can be a good complementary method to combine with other approaches and gain sensitivity to systematics and robustness to the science.

\section*{Acknowledgements}
AP, FS and JB acknowledge support from a European Research Council Starting Grant (LENA-678282). This work has been supported by MINECO  grants AYA2015-71825, PGC2018-102021-B-100  and  Juan de la Cierva fellowship, LACEGAL and EWC Marie Sklodowska-Curie grant No 734374 and no.776247 with ERDF funds from the EU  Horizon 2020 Programme. IEEC is partially funded by the CERCA program of the Generalitat de Catalunya.


\bibliographystyle{aa}
\bibliography{biblist}

\appendix

\section{Details of the training}\label{sec:training}

\begin{figure}
\centering
\includegraphics[width=.98\linewidth]{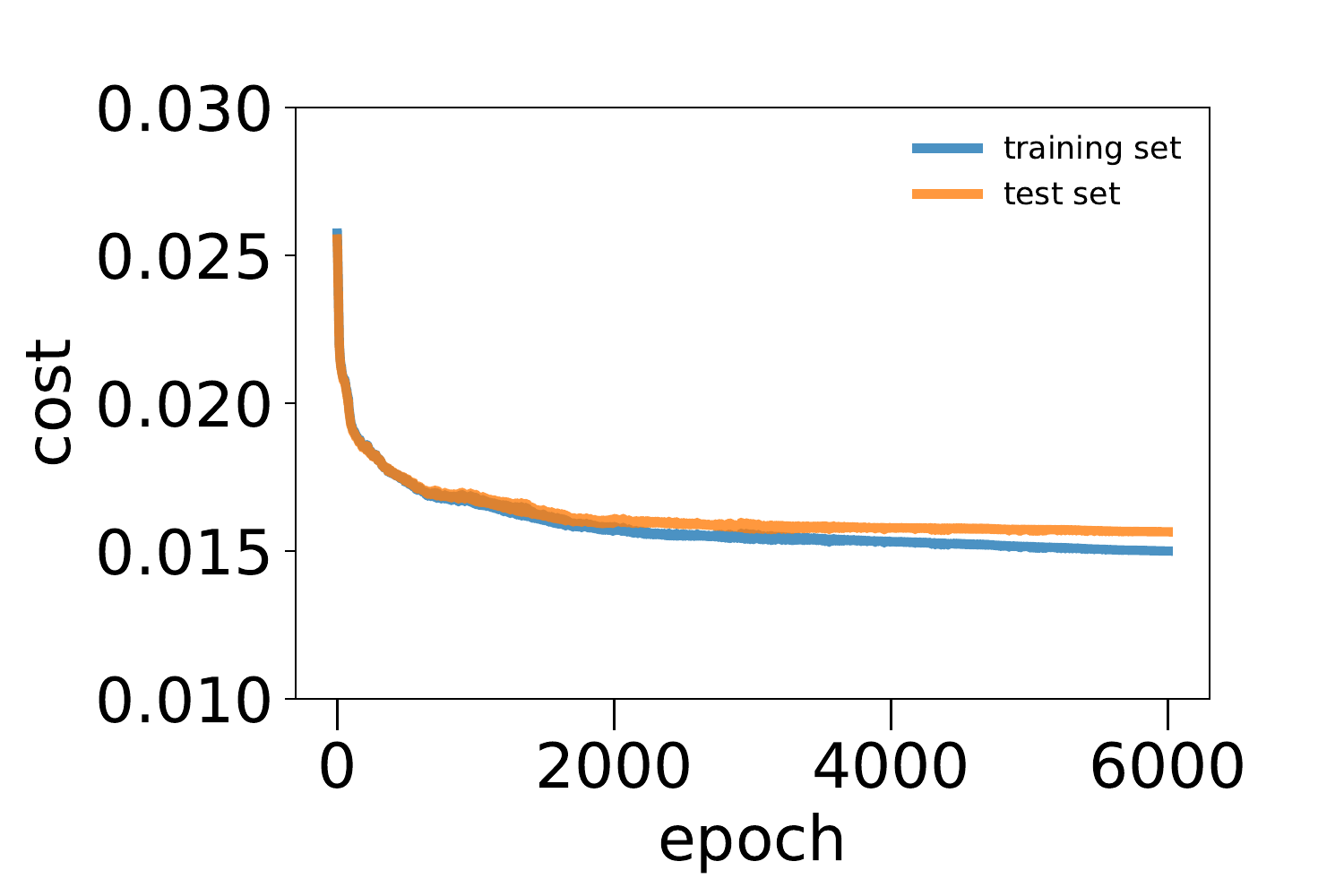}
\caption{Evolution of the cost function for the training and the test sets during the training on the fiducial implementation of NNSC.}
\label{fig:costs}
\end{figure}

\begin{figure*}
\centering
\includegraphics[width=.48\linewidth]{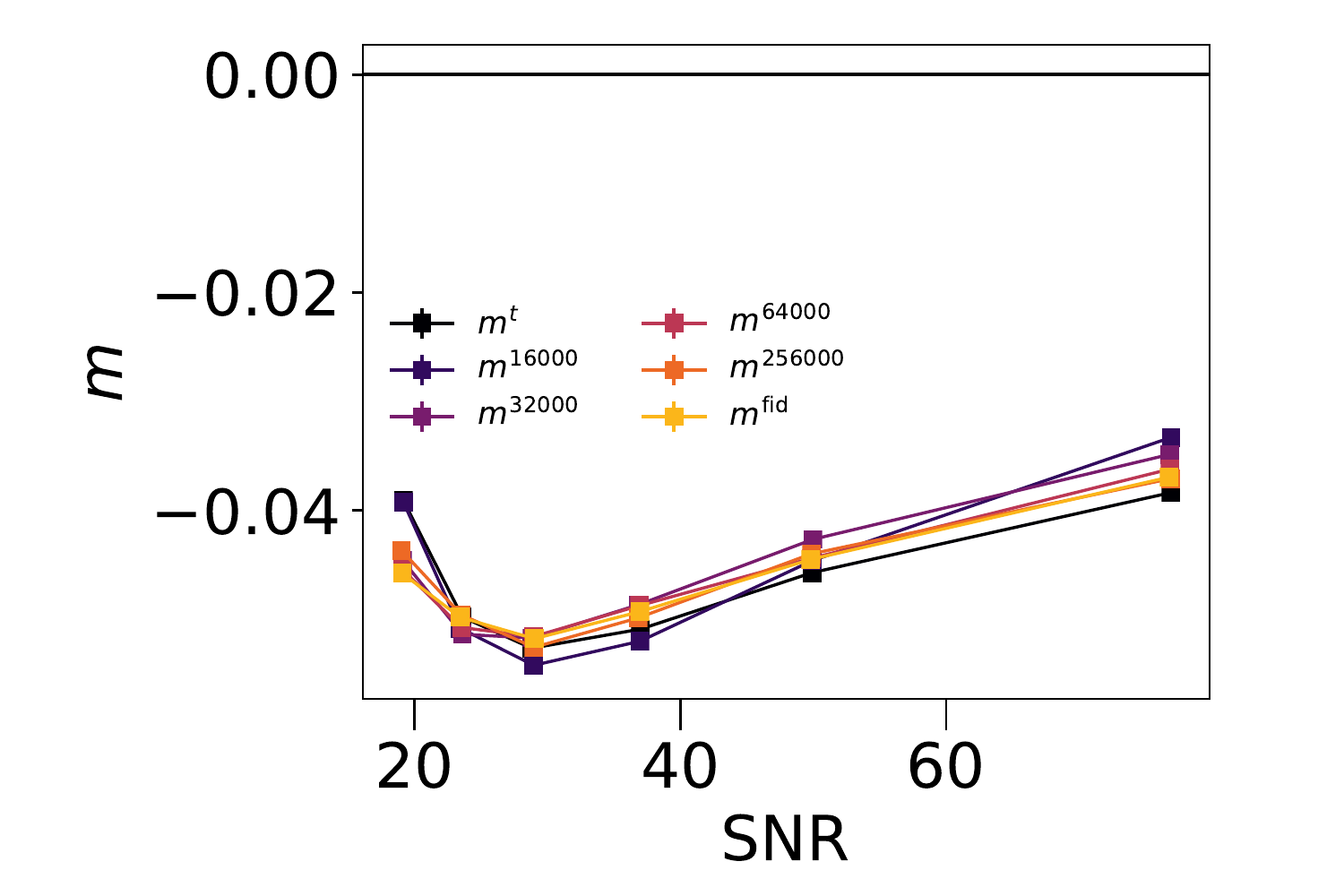}
\includegraphics[width=.48\linewidth]{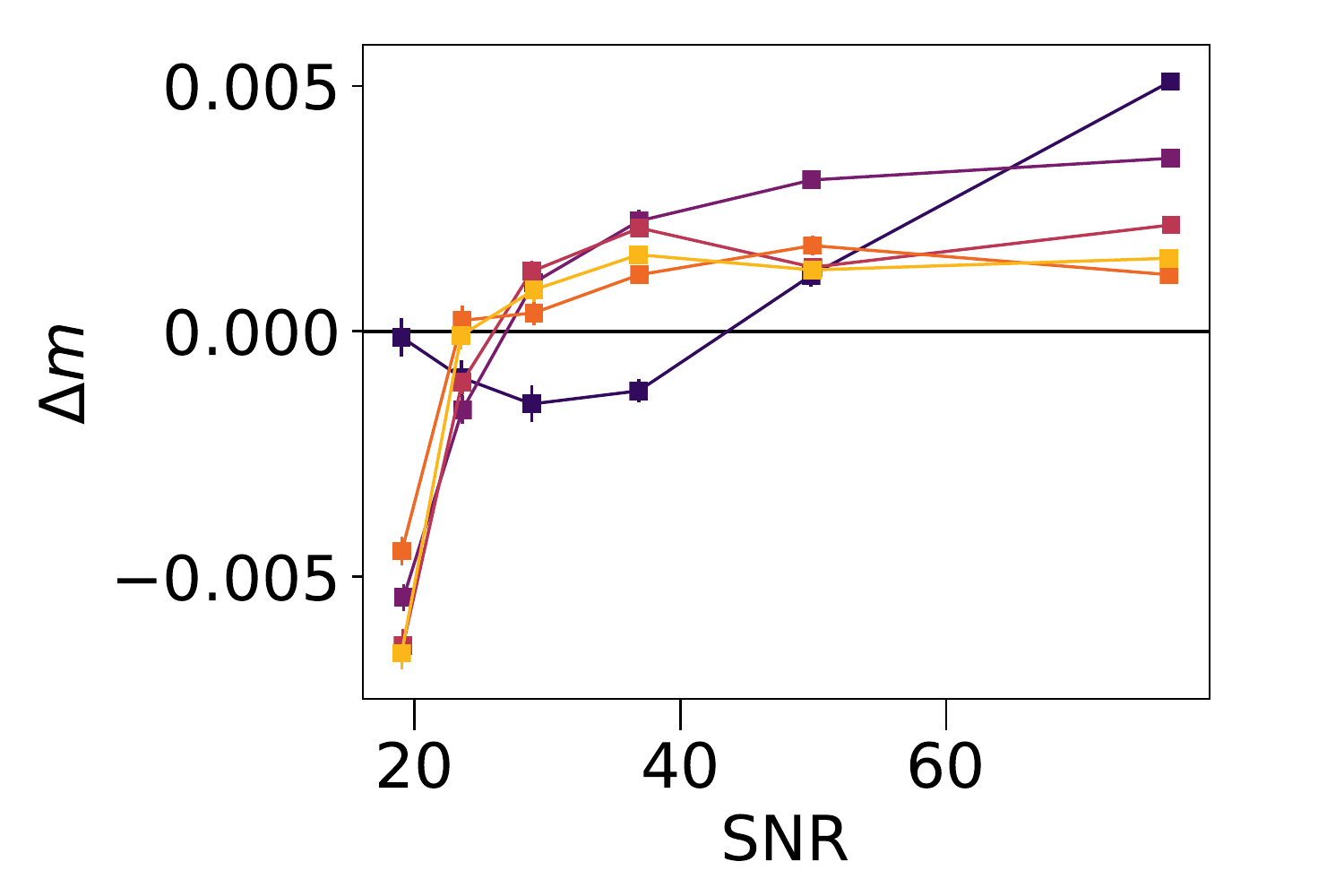}
\includegraphics[width=.48\linewidth]{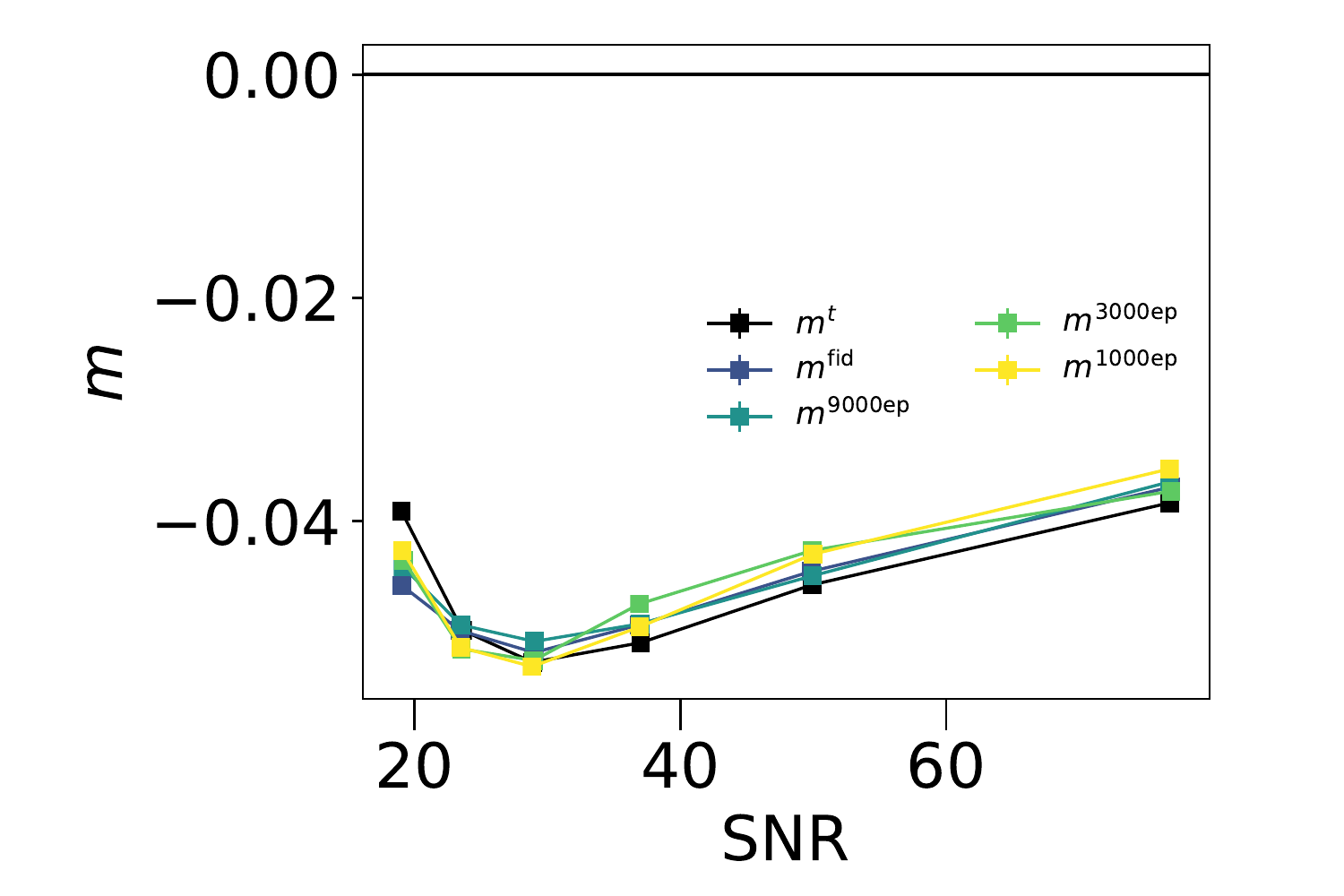}
\includegraphics[width=.48\linewidth]{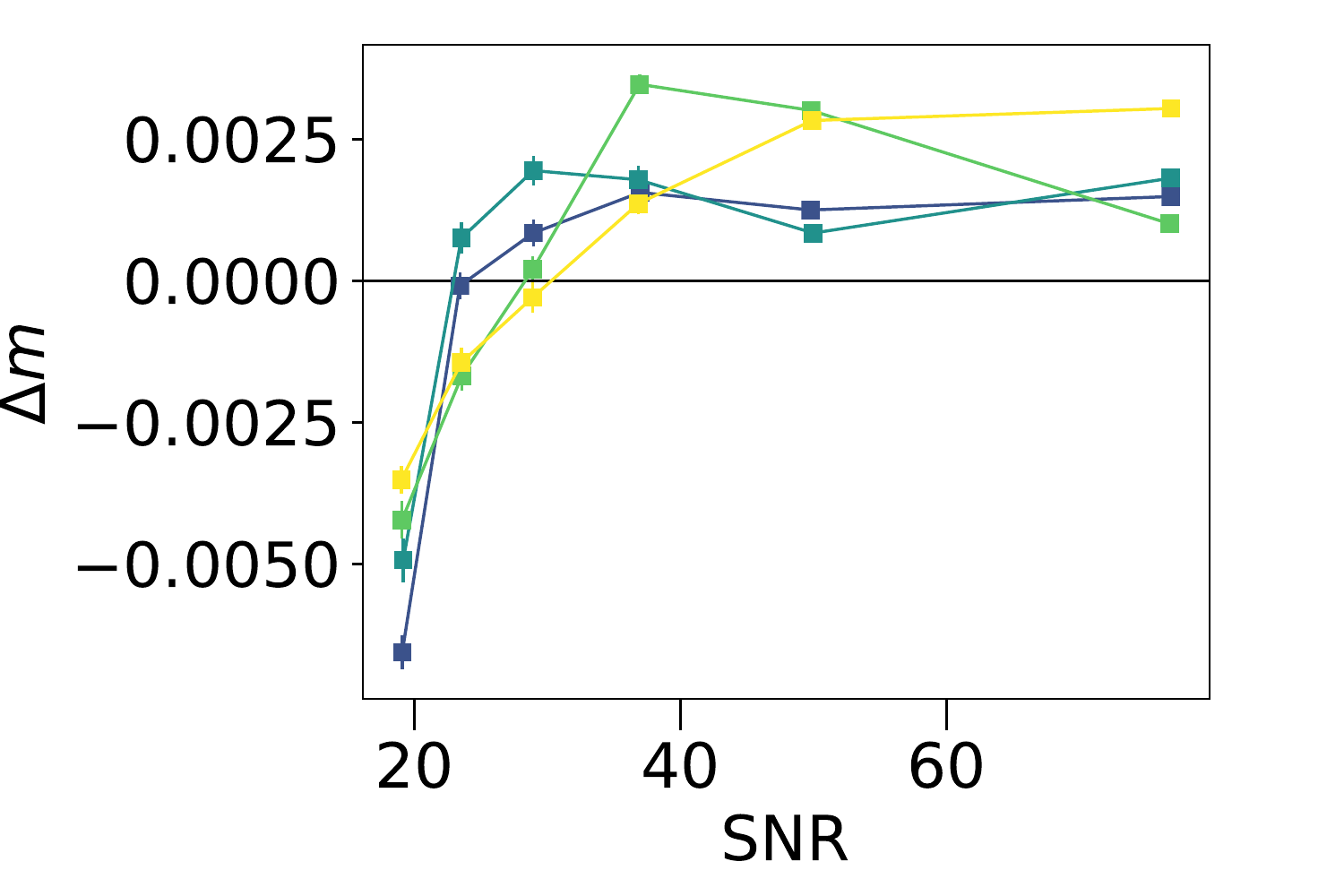}
\includegraphics[width=.48\linewidth]{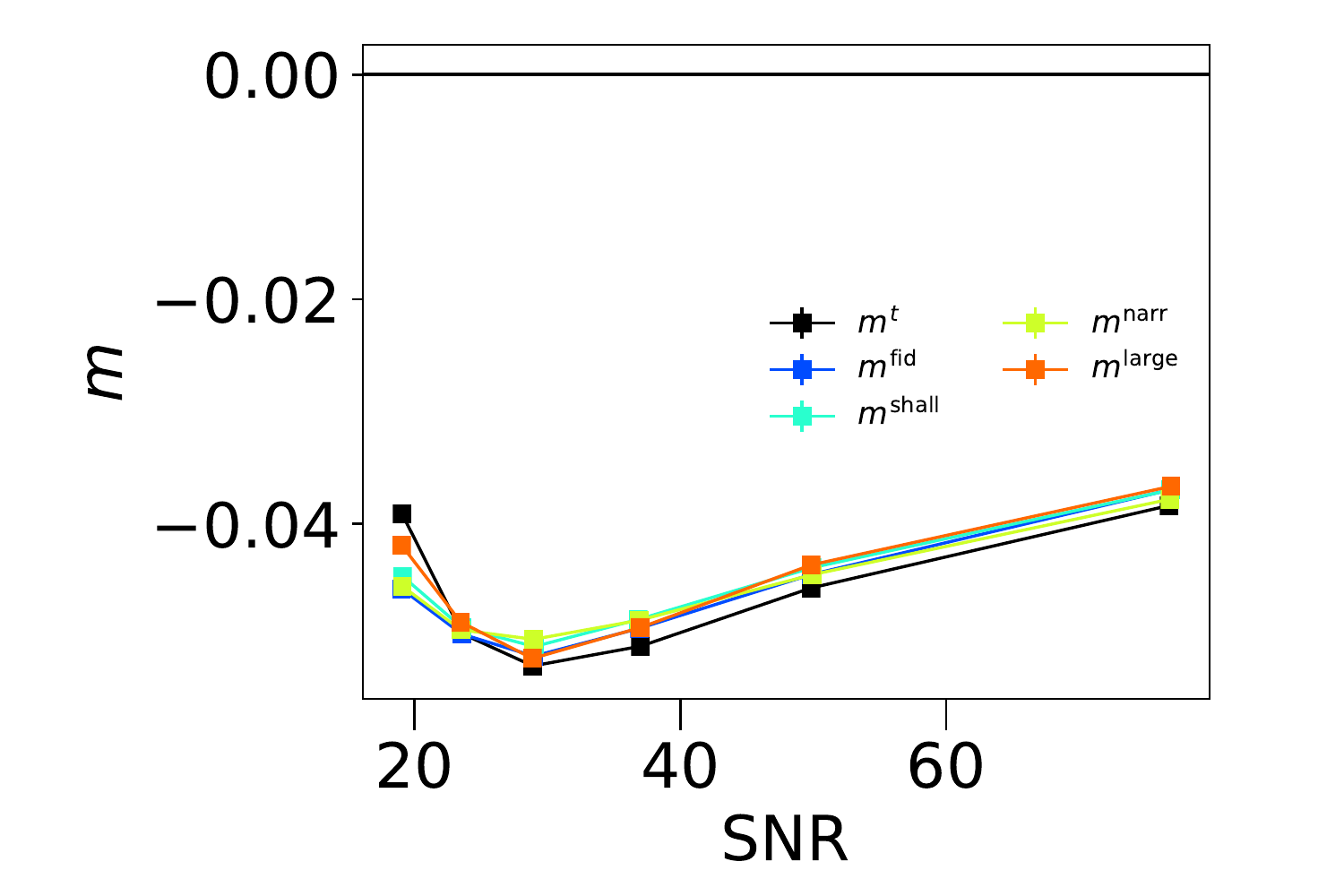}
\includegraphics[width=.48\linewidth]{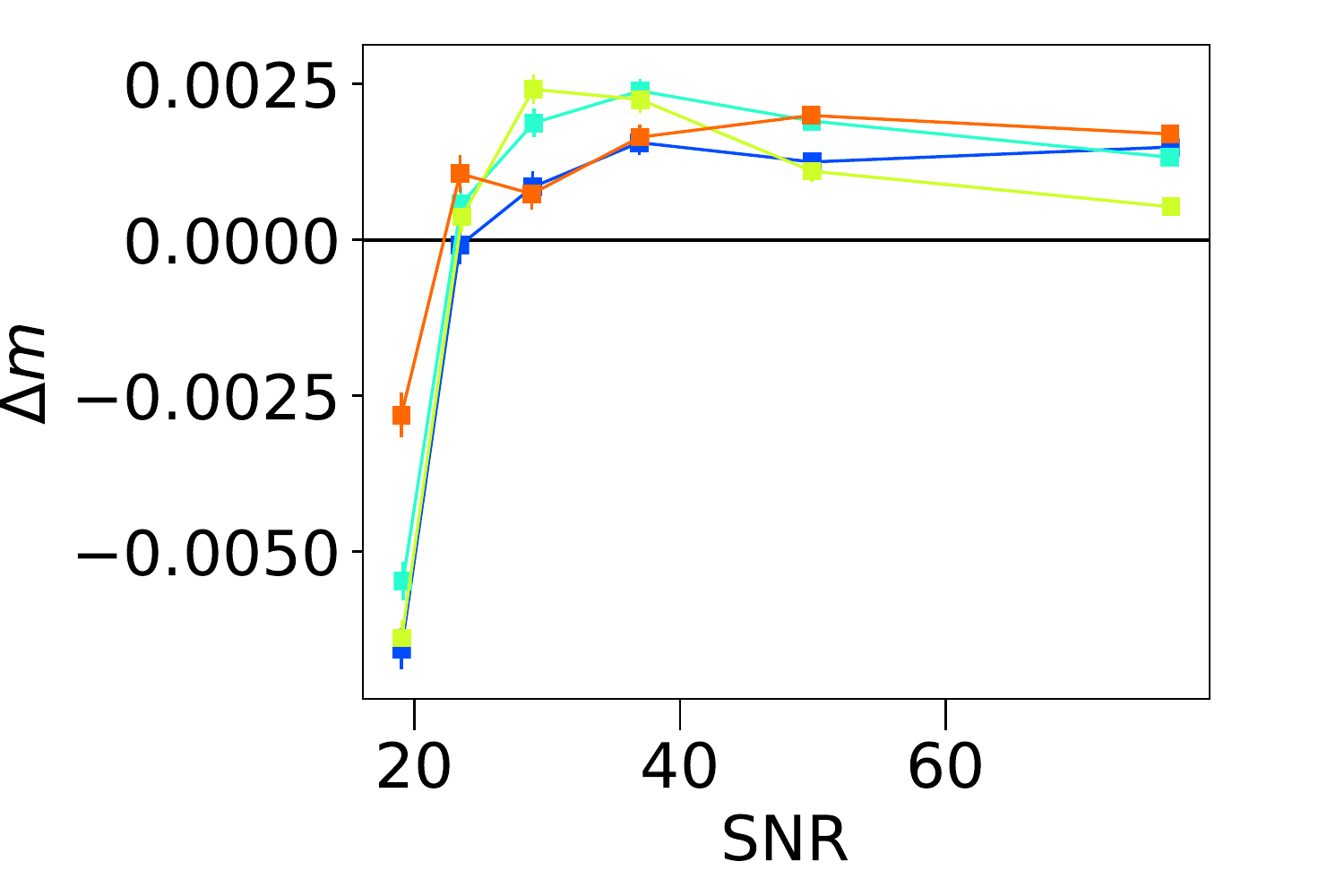}
\caption{Comparison of estimated multiplicative shear bias $m_1$ (left) and its error $\Delta m_1$ (right) as a function of S/N for different ML hyper-parameters. The top panels show the performance for different numbers of objects used in the training, from $10,000$ to $400,000$. The middle panels show the performances for the chosen architecture for different number of epochs, from 1,000 to 9,000. The bottom panels show the performance for different architectures. The chosen one, shown in blue, corresponds to $\text{four}$ hidden layers of 30 nodes each. The shallow architecture, shown in green, has only $\text{two}$ hidden layers of 30 nodes. The narrow architecture, shown in orange, corresponds to $\text{four}$ hidden layers of 30, 20, 10, and 10 nodes.}
\label{fig:1d_m_comp}
\end{figure*}

\begin{figure*}
\centering
\includegraphics[width=.48\linewidth]{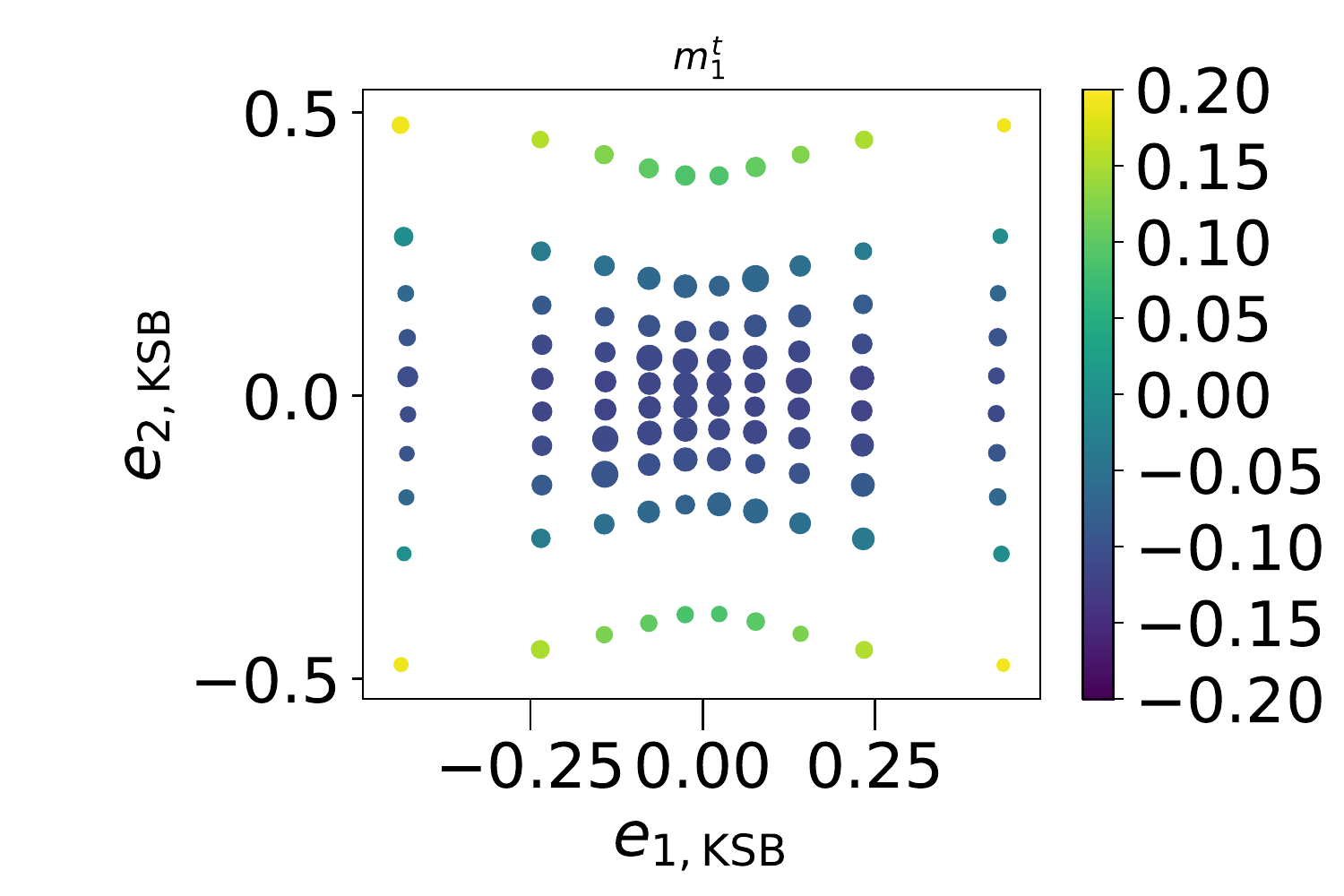}
\includegraphics[width=.48\linewidth]{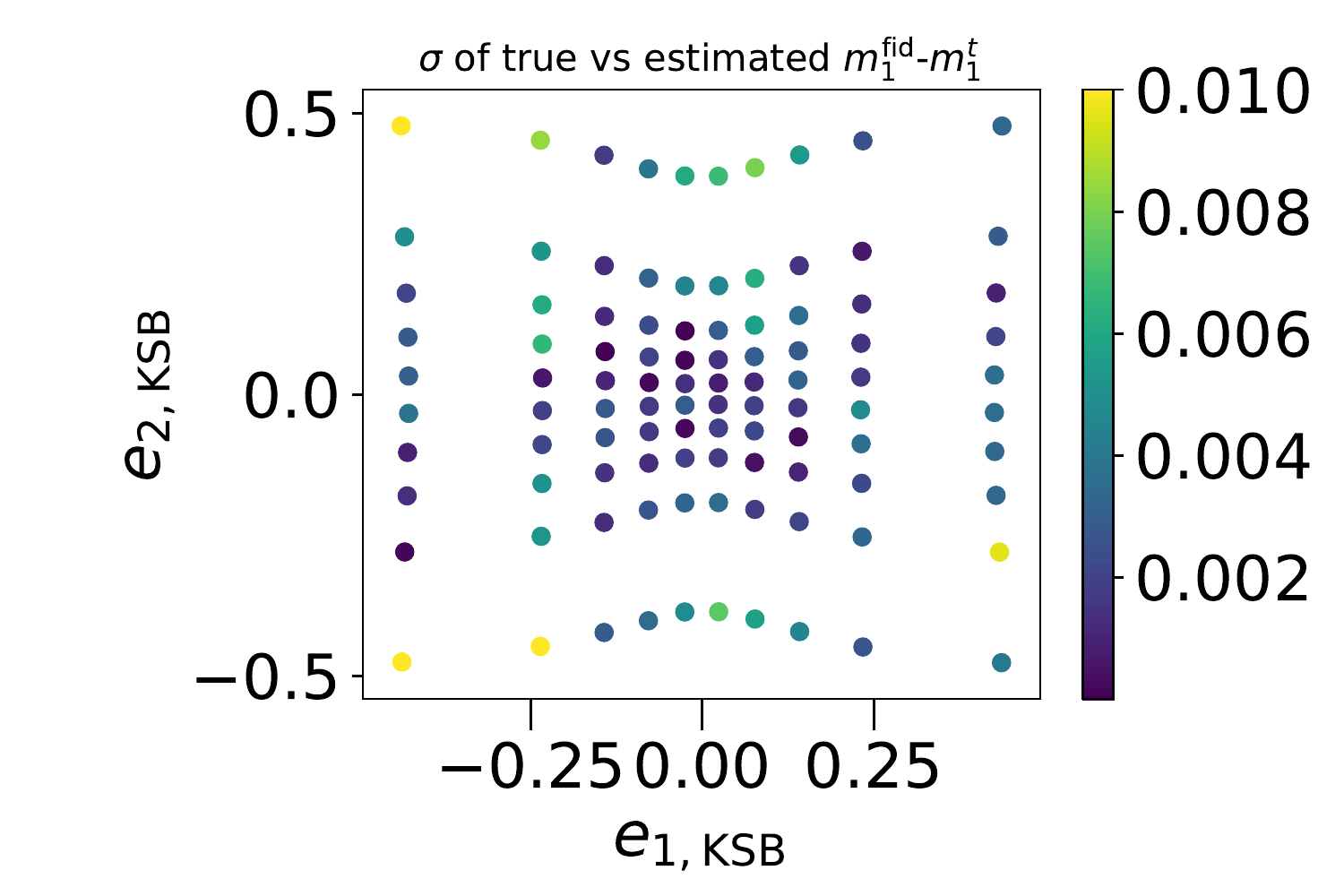}
\includegraphics[width=.48\linewidth]{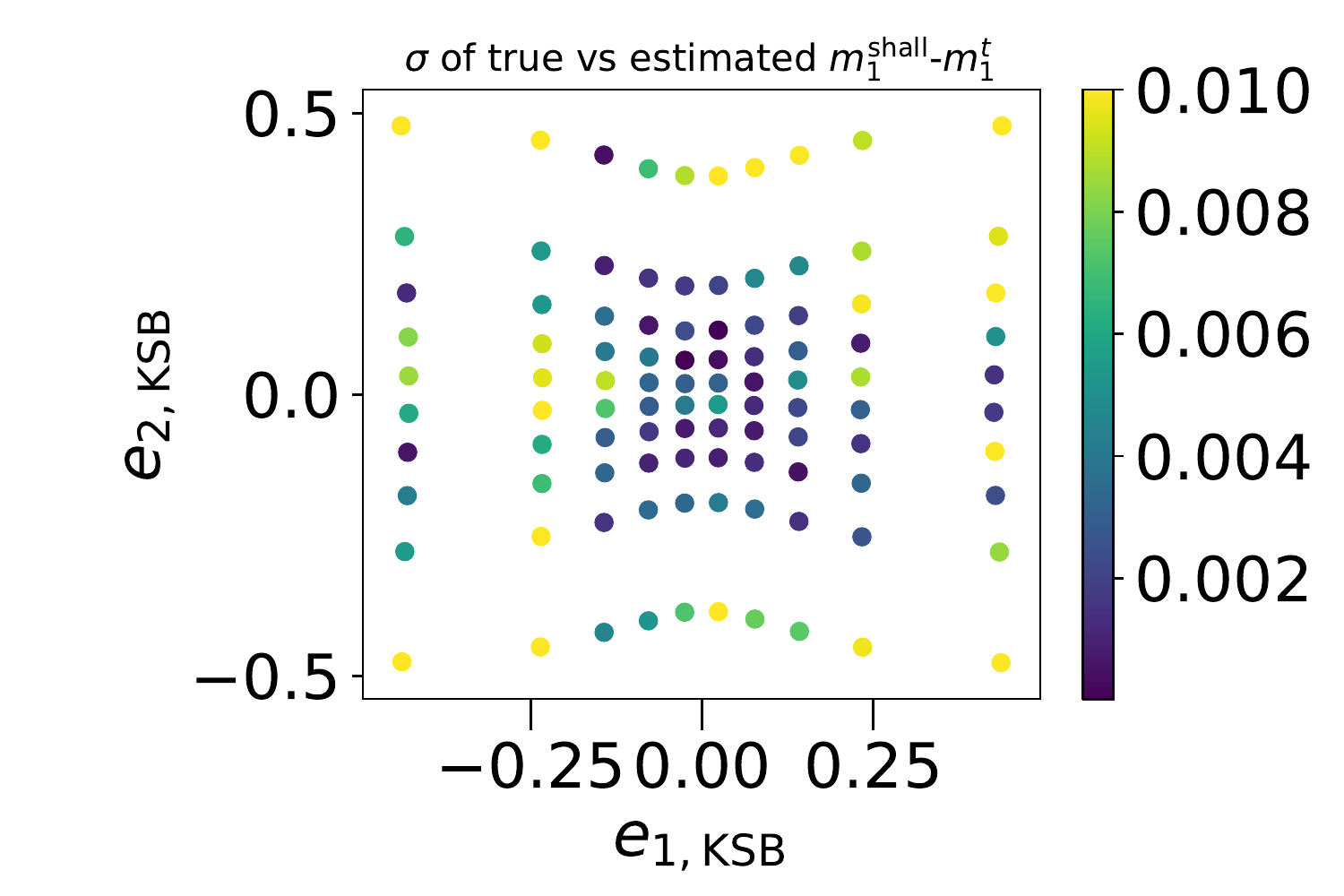}
\includegraphics[width=.48\linewidth]{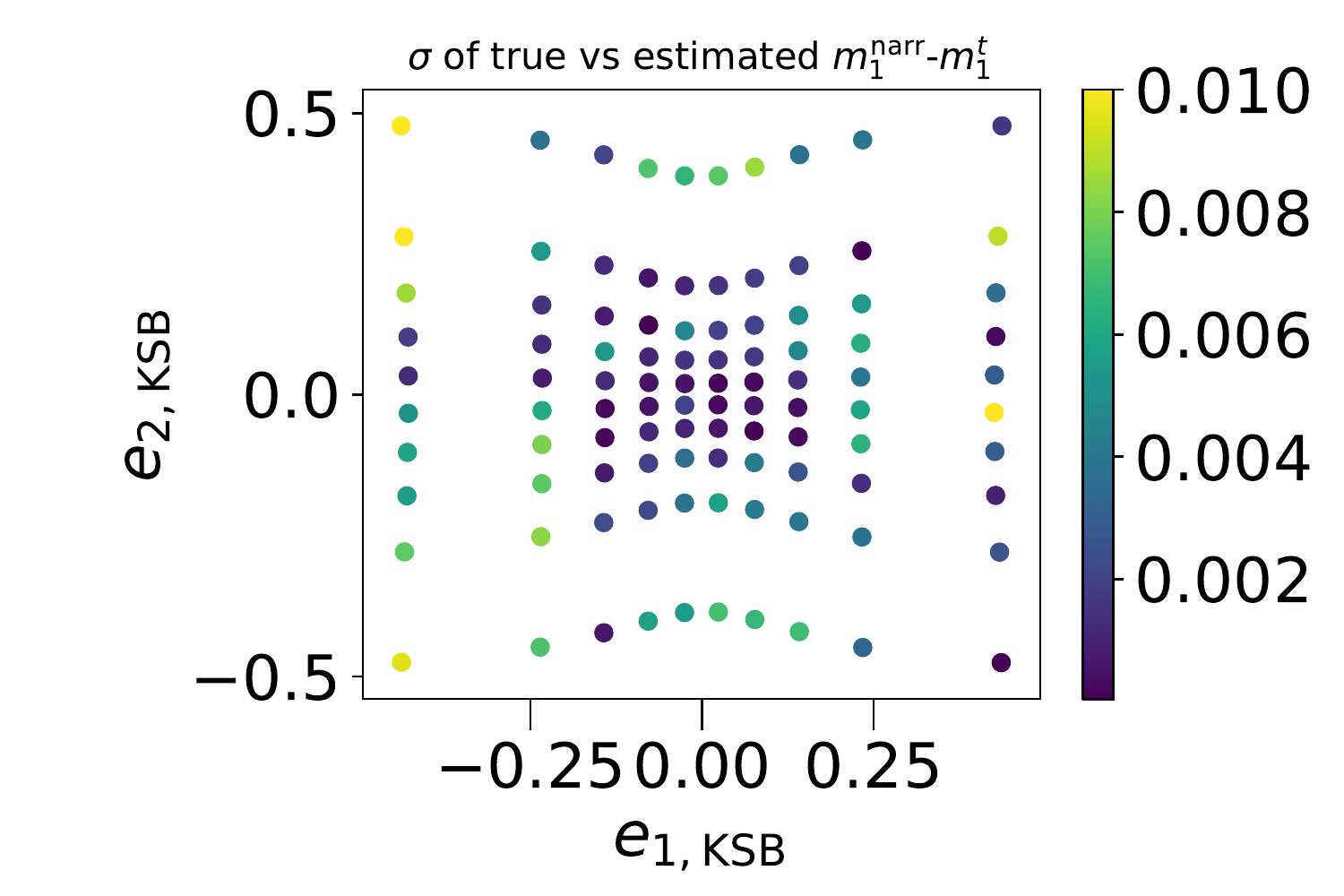}
\caption{Comparison of estimated multiplicative shear bias $m_1$ as a function of galaxy measured ellipticities. The top left panel shows the true values of $m_1$, and the remaining panels show the difference between the estimated and the true for the fiducial architecture (top right), the shallow architecture (bottom left), and the narrow one (bottom right).}
\label{fig:2d_m_comp_e1e2}
\end{figure*}

In this section we describe the training procedure and the architecture, cost function, and hyper-parameters we implemented for the results. This configuration was chosen according to the performances found during the optimisation process. Even if these optimisation parameters could be further studied and optimised, we already obtained competitive results with the procedure and the optimisation tests described below.

Our configuration consists of a DNN with four hidden layers of $30$ units per layer. The input consists of the $27$ measured properties as described before, and the output consists on the $\text{four}$ shear bias components.

We first applied a whitening with principal component analysis (PCA) to the input data in order to decorrelate the $27$ properties, and we normalised them to be between $0$ and $1$. This procedure aims at avoiding some properties or information to dominate their contribution due to their value ranges or correlations.

Here we recall the cost function $C,$ which minimises the $\chi^2$ on the estimated shear biases,
\begin{equation}
C = \frac{1}{b_\textrm{s}} \sum_{i=0}^{b_\textrm{s}} \sum_{\alpha = 1}^2 (m_{i,\alpha}^\textrm{t} - m_{i,\alpha}^\textrm{e})^2 + (c_{i,\alpha}^\textrm{t} - c_{i,\alpha}^\textrm{e})^2,
\label{eq:cost}
\end{equation}
where $m_\alpha^\textrm{t}$, $c_\alpha^\textrm{t}$ and $m_\alpha^\textrm{e}$, $c_\alpha^\textrm{e}$ are the true and estimated $\alpha$th component shear multiplicative and additive bias, respectively, and $b_\textrm{s}$ is the batch size. Given that usually $m_{i,\alpha} >> c_{i,\alpha}$, the contribution of $m_{i,\alpha}$ in the current approach dominates the minimisation process, but additional weights can be applied to some biases when the performance of the estimation of these biases is to be differently. Here we include the four bias components in the cost function to estimate them simultaneously, but separate trainings for each of the components can also be made. Separate trainings for each of the components might allow using simpler architecture and faster learning, but it would miss the correlation between the components in the learned model.

We constrained the hyper-parameters of the algorithm (contamination level, number of training objects, number of epochs, batch size, learning rate, and learning decay rate) by analyzing the performance and convergence in a wide hyper-parameter space, choosing the final hyperparameter set from the best cases that we found by avoiding overfitting according to the cost function values of the training and test sets (see Fig.~\ref{fig:costs} for the evolution of the cost function during the training of the fiducial example). For this, we evaluated the cost function in the training and the test sets to ensure that, on one hand, the costs converge during the training, and on the other hand, that the cost in the training set is not significantly lower than the cost in the test set, which would be a symptom of overfitting. In some cases, the cost in the test set was found to be lower than in the training set. These cases were also discarded for this accidental overfitting.

In Fig. \ref{fig:1d_m_comp} we show the performance of the shear bias estimates for different hyper-parameters of the ML optimisation. The left panels show the shear bias dependence as a function of S/N for the true multiplicative bias and the estimated biases. The right panels show the difference between the estimated and true bias as a function of S/N.

In the top panel we show the performance for a different number of objects used in the training set, going from $16,000$ to $256,000$ objects (the fiducial case was computed with $128,000$ objects). More objects improve the performance, which reaches some plateau for more than $50,000$ objects. This is a very small number of objects compared with most of the shear calibration approaches \citep{Zuntz2017,Tewes2018,Kannawadi2019}.

The middle panels show the performance as a function of the number of epochs used in the training (with $6,000$ epochs for the fiducial case). In this case, the training converges for more than $6000$ epochs. For our final case we used $6000$ epochs, which took about $15$ CPU hours of training, but we note that similar results are obtained with longer trainings. This shows that our method has a very fast training compared to other ML approaches.

In the bottom panels we compare the performance using different architectures. Our final case uses four hidden layers with 30 nodes in each layer, but here we compare it with a case of a narrower architecture (with four hidden layers with 30, 20, 10, and 10 nodes), a shallow one (with two hidden layers of 90 and 10 units), and a larger one (with four hidden layers of 50 nodes each). The performances as a function of S/N are very similar, with no obvious conclusion about which architecture is giving better predictions. However, Fig.\ref{fig:2d_m_comp_e1e2} shows the interest of using a wider and deeper architecture. In the top left panel, we show the true $m_1$ as a function of two properties (the two ellipticity components measured by \ksb). The other panels show the difference between the estimated and the true $m_1$ for the large architecture (top right), the shallow (bottom left), and the narrow (bottom right) architectures. The narrow and shallow architectures are not able to capture the entire complexity of the 2D dependence as efficiently as our fiducial architecture. This is proof that a wide and deep neural network allows us to better capture the complexity of the system. Although not shown here, the performances of the larger architecture are very similar to those of the fiducial one, and the performance is poorer for the same training time. This indicates that the largest architecture does not help improve the performance and loses efficiency of the training, and for this reason, we kept the four hidden layers of 30 nodes as the fiducial model for the purpose of this paper.

For the remaining hyper-parameters we did not find a strong dependence on the batch size, showing good performance between $16$ and $256$ ($32$ were finally chosen), and we found that adding no noise contamination to the input data was optimal for the performance. About the activation function of the layers, we used a leaky ReLU function for the nodes. We found similar performances using $\tanh{}$ functions or combinations of both, but with a significantly slower convergence.  The results shown in the paper for the chosen model were obtained with $\text{about } 15$ CPU hours.

\section{Metacalibration}
\label{sec:metacal}

The calibration method \metacal\ has been presented in \cite{Huff2017} and \cite{Sheldon2017}, with a publicly available implementation\footnote{https://github.com/esheldon/ngmix/wiki/Metacalibration}. It has been tested on simulations and applied to the Dark Energy Survey (DES) Y1 data \citep{Zuntz2017}, showing a very good performance. We used this as a reference recent calibration method for comparison with our new approach. We briefly describe the idea of this method and refer to \cite{Huff2017} and \cite{Sheldon2017} for more details.

\metacal\ is based on measuring the shear response of the individual galaxy images without any need of image simulations. To do so, \metacal\ uses the original real image to generate sheared versions of it. With these sheared versions, the shear response is obtained using Eq. \ref{eq:shear_response} as in our method. The shear calibration is then applied to the data using the mean of these individual shear responses and their propagation through the statistics of interest.

To generate the sheared versions of the original image, \metacal\ first deconvolves the image with the PSF (which is assumed to be perfectly known). Then the shear distortion is applied, and the image is reconvolved with a slightly higher PSF. As this process induces correlated sheared noise, a noise image following the same procedure but with opposite sign shear is also added to reduce the effect of this correlated shear on the shear response. The final noise realisations can be significantly different than the original one, which can have an effect on the shear response. Because of this, a non-sheared new image is also generated with the same process. On this new image, shear is measured and calibrated for science analyses. In addition, the additive bias can be measured using Eq. 18 from \cite{Huff2017}, and the shear response coming from selection biases can be estimated \citep{Sheldon2017}.

\metacal\ has the advantage that no image simulations for the calibration are required (although its performance can only be tested in simulations). On the other hand, the method depends on the numerical processes involving deconvolution, reconvolution, and the treatment of noise.

\metacal\ allows for different implementations regarding the characterisation of the PSF, including a Gaussian parameter fitting of the PSF (so that a combination of Gaussian profiles is used as the PSF), a symmetrisation (where three different rotations of the PSF are stacked to avoid a contribution of its ellipticity), and using the true PSF directly. We used the true PSF, but we found very similar results using the symmetrisation. Moreover, \metacal\ can be applied for any shape measurement algorithm for which the method calibrates its shear bias. We used \metacal\ to calibrate the \ksb\ mesurements from the software \shapelens.

\end{document}